\documentstyle[preprint,eqsecnum,prd,aps]{revtex}
\input{psfig}
\begin{document} 
\renewcommand{\thefootnote}{\fnsymbol{footnote}}
\setcounter{equation}{0}
\newcommand{\beq}{\begin{equation}}
\newcommand{\eeq}{\end{equation}}
\newcommand{\beqa}{\begin{eqnarray}}
\newcommand{\eeqa}{\end{eqnarray}}
\input{epsf}
\pagestyle{plain}

\preprint{
\vbox{
\halign{&##\hfil\cr
	& ANL-HEP-PR-95-84\cr
        & IFT-UFL-95-28 \cr}}
}
\title{
Polarized and Unpolarized Double Prompt Photon Production
in Next-to-Leading Order QCD}
\author{Claudio Corian\`{o}$^{a,}$ and L. E. Gordon$^b$}
\address{
$^a$Institute for Fundamental Theory,
Physics Department,\\
 University of Florida at Gainesville, 32611, FL, USA\\
$^b$ High Energy Physics Division, Argonne National Laboratory,
	Argonne, IL 60439, USA }
\maketitle
\begin{abstract} 
We calculate $O(\alpha_s)$ corrections to inclusive and isolated double
prompt photon production, both for the unpolarized case, and for
longitudinal polarization of the incoming hadrons. The calculation is
performed using purely analytical techniques for the inclusive case, and
a combination of analytical and Monte Carlo techniques to perform the
phase space integration in the isolated case. A brief
phenomenological study is made of the process $pp\rightarrow \gamma
\gamma X$ at CMS energies appropriate for the RHIC heavy ion collider.
\end{abstract}
\vspace{0.2in}

\narrowtext
\section{Introduction}
Among the various and still unexplained aspects of QCD, the study of the 
spin structure of the nucleon has gathered considerable work in 
recent years \cite{EMC}, suggesting that a simple parton model interpretation 
of the phenomenon is far too simplified to completely describe the distribution 
of spin inside the nucleons. The original EMC data suggested that 
quark spin accounts for only a small fraction
of the nucleon spin. Since then the possibility that gluons may carry a 
significant fraction of the nucleon's spin has been advanced and 
investigated theoretically (see \cite{set1}  and Refs. therein). But
other alternative explanations, such as the possibility that the proton
may have a large negative strange sea polarization have also been
suggested, along with various scenarios which use a combination of the
two extremes.    

Recently, new experimental results on the polarized proton structure function 
$g^p_1$ have become avaliable \cite{set2}. Most of the information collected 
so far from the phenomenological side, which however still leaves an 
appreciable disagreement with naive parton model expectations, is based on 
studies of the hadronic tensor in the kinematic region of Deep Inelastic 
Scattering (DIS), through its structure functions $g^p_1(x, Q^2)$ ($x$ is the 
Bjorken variable), primarily at large $Q^2$ and finite $x$. 
This discrepancy, referred to by some as ``the spin crisis'', indicates that 
sum rule predictions \cite{EJ}, based on the constituent quark model
picture of the nucleon, need to be amended, since it is only in the DIS 
limit that the structure functions - which parametrize the hadronic tensor - 
scale to quark distributions which are functions of $x$ only.

At finite $Q^2$, DIS sum rules are violated by quark-gluon interactions, 
usually parameterized by higher twist effects in the Operator Product 
Expansion. The contribution coming from the axial anomaly
\cite{set1} (see also \cite{ans} and Refs. therein), in particular, 
has been shown to induce a cancellation between quark and gluon contributions 
to the first moment of the polarized structure function, $g^p_1(x)$. 
 
Although the anomalous contribution is a radiative effect $(O(\alpha_s))$, 
renormalization group arguments and the $\log (Q^2)$ increase of the 
nucleon spin suggest that such contribution 
 (which is $\alpha_s (Q^2) \Delta G$) remains finite even in the $Q^2\to 
\infty$ limit in which the O.P.E. is applicable. The ``spin crisis'' is 
therefore solved by the observation that the EMC result is $not$ 
a measure of the quark spin, according to the naive parton model picture. It 
only measures the combination 
$$   
{1\over 9}\left(\Delta\Sigma - {3\alpha_s\over 2 \pi}\Delta\,G\right)
$$
between the quark $\Delta\Sigma$ and the gluon contributions to the total 
spin. This leaves open the question as to the actual size of $\Delta G$,
i.e. whether it is large or small. This question can only be settled
conclusively by experiments which are capable of directly probing
$\Delta G$. 

Given the universality of the structure functions within QCD, it is expected, 
in a few years from now, that the information gained from DIS polarized 
scattering will be supported and supplemented by new results from polarized 
proton-proton collisions at the BNL heavy ion collider RHIC in forthcoming 
experiments. It is therefore important and interesting to see how various  
processes at large $p_T$ are affected -in the polarized case- 
in their behaviour when next-to-leading (NLO) order corrections are included.  
It is also interesting to see how the asymmetry between the unpolarized and the 
polarized cross sections behaves when a reduced scale dependence is present, 
due to firmer NLO  perturbative calculations. From the theoretical side, 
further motivation to proceed with these studies comes from the fact that the 
splitting functions for a complete NLO $Q^2$-evolution of the spin structure 
function $g_1$ have been recently obtained \cite{vn}, and new parametrizations
in NLO will soon be available for the first time. Many
processes have been suggested for study at RHIC \cite{robinett}, but only a
few so far have been calculated at NLO \cite{nlo1,nlo2}.

In this work we present a next-to-leading order study of the process 
$p p\rightarrow\gamma \gamma X$, which has been calculated in the past 
\cite{ael}  for unpolarized photons at NLO. We repeat both these calculations 
and extend them to the polarized case, which to our knowledge has only
been studied in LO \cite{rd}, with inclusion of the higher order 
($O(\alpha_s^2)$) process
$gg\rightarrow \gamma \gamma$. There it was found that the cross section should
be large enough to measure, and should have some sensitivity to the
polarized parton distributions but that the asymmetries are not very large
except in regions where sea quark spin dependent processes dominate.
This meant that the process would not be useful for extracting
information on $\Delta G$. But they also made the point that a full NLO
study including all relevant contributions would be necessary to confirm
these conclusions, since cancellations between the various processes may
occur, and the effect of the corrections were completely unknown. 
For example, they could not include
contributions for the higher order process $qg\rightarrow\gamma \gamma
q$, which are known to be quite significant, and can even be larger than
the $q\bar{q}$ scattering process in proton-proton collision
processes, as opposed to proton-anti-proton collision processes.
Also, in addition to the possibility that double
prompt photon production may yield important information on the
polarized proton structure function, particularly the possibility that a
large fraction of proton spin is carried by gluons, or a large
negatively polarized strange sea, the process is also important because it is a
background to Higgs decay. A full NLO analysis using more modern
polarized parton distribution is now necessary to either support or
disagree with the conclusions of Ref.\cite{rd}. In this study we are
not yet able to use parton distributions evolved in NLO as these are not
yet ready, but in a future study we hope to include them
\cite{cg}. We can nevertheless extend that of
Ref.\cite{rd}, by including all O($\alpha_s$) higher order corrections to the
hard scattering process and correctly taking isolation effects into
account. 

The organization of the paper is as follows. In section II we describe the 
methods involved in the calculation and the regularization prescriptions 
adopted for the inclusive case. The main contribution of this paper is the 
calculation of the polarized cross section, but since the unpolarized cross 
section has been
presented before, we discuss this first in each case and then extend
the arguments to the polarized case. In section III we outline the
combined analytic/Monte Carlo method of calculating the isolated cross
section. In section IV we present and discuss some numerical results, and
in section V we draw some conclusions.

\section{Inclusive Double Photon Production}

The inclusive cross section we calculate is 
\begin{displaymath}
\frac{d\sigma}{dk_{T1}^{\gamma}dy^{\gamma}_1dz}
\end{displaymath}
This cross section is differential in the transverse momentum,
$k_{T1}^{\gamma}$,
and rapidity of one of the photons (referred to as the trigger photon) 
plus the variable $z$ defined by
\begin{equation}
z=-\frac{k_{T1}^{\gamma}.k_{T2}^{\gamma}}{|k_{T1}^{\gamma}|^2}.
\end{equation}
$z$ thus contains information about the transverse momentum of the second
photon, but not its rapidity. The cross section is thus limited in that
experimental cuts on the rapidity of the second photon and isolation
cuts on either photon cannot be implemented. It is nevertheless still
worth calculating, since it provides a check on the more versatile
calculation using combined analytic and Monte Carlo techniques which we 
describe in section III.  

\subsection{LO Contributions}

In LO $O(\alpha_{em}^2)$, the contributions to double prompt photon 
production are from $q\bar{q}$
annihilation to two photons, and the various single and double fragmentation
contributions, some of which are shown in Fig.1. From Eq.(2.1) we can see
that the $q\bar{q}$ annihilation process in LO is proportional to
$\delta(1-z)$, as the $k_T's$ of the two photons must balance. The hard
subprocess cross section is
\begin{equation}
\frac{d\hat{\sigma}}{dvdwdz}(q\bar{q}\rightarrow \gamma \gamma)=
\frac{2\pi\alpha_{em}^2}{3\hat{s}}e_q^4\frac{1-2v+2v^2}
{v(1-v)} \delta(1-z)\delta(1-w)
\end{equation}
where 
\begin{eqnarray}
v&=&1+\frac{{t}}{{s}}\nonumber \\
w&=&\frac{-{u}}{{s}+{t}},
\label{vw}
\end{eqnarray}
$\alpha_{em}$ is the electromagnetic coupling constant and $e_q$ denotes
the quark charge. The usual Mandelstam invariants are defined in terms
of the momenta of the two incoming hadrons $P_A$ and $P_B$, the momentum
fractions of the initial partons $x_1$ and $x_2$, and the momentum of
the trigger photon $k_1$ via,
\begin{eqnarray}
{s}&=&(x_1P_A+x_2P_B)^2=x_1x_2 S\nonumber \\
{t}&=&(x_1P_A-k_1)^2 \nonumber \\
{u}&=&(x_2P_B-k_1)^2
\end{eqnarray}
where $\sqrt{S}$ is the center-of-mass energy in the hadronic system.

In this study we do not include contributions where both photons are
produced by fragmentation off a final state parton as these are expected
to give a very small contribution, especially after isolation effects
are considered. We do include process where one of the photons is
produced directly and the other by fragmentation, referred to as the
single fragmentation contributions (Fig.1b). These processes contribute
to $O(\alpha_{em}^2)$, although the hard subprocess is
$O(\alpha_s\alpha_{em})$, since the fragmentation function of parton $i$
into a photon, $D_{\gamma/i}(z,Q^2)$ is $O(\alpha_s/\alpha_{em})$. In a
fully consistent NLO calculation we should also include the
O($\alpha_s$) corrections to these fragmentation processes, but we are not 
yet in a position to provide these.  

The full expression for the physical $q\bar{q}$ annihilation cross section 
in LO is
\begin{equation}
\frac{d\sigma}{dk_{T1}dy_1dz}=\frac{2\pi
k_{T1}}{\pi S}\int^V_{VW}\frac{dv}{1-v}f^A_q(x_1,M^2)f^B_{\bar{q}}(x_2,M^2)
\frac{d\hat{\sigma}}{dvdwdz}(q\bar{q}\rightarrow\gamma\gamma)+(q\leftrightarrow
\bar{q}),
\end{equation}
where $V$ and $W$ are defined similarly to $v$ and $w$ in Eq.(2.3) except
now in the hadronic system, and $f^A_i(x,M^2)$ is the parton distribution
function for parton of type $i$ in hadron $A$ as a function of $x$, the
momentum fraction, and the scale, $M^2$. 

There are two types of single fragmentation contributions to the process. Those
where the trigger photon $\gamma_1$ is produced by fragmentation and
those where $\gamma_2$ is. These contribute in different regions of $z$,
the former in the region $z\geq 1$ and the latter in the region $z\leq
1$. In terms of the hard subprocess cross sections for the processes
$q\bar{q}\rightarrow \gamma g$ and $qg\rightarrow\gamma q$ which can be
found, for example in Ref.\cite{ael},
these contributions are given by the expressions;
\begin{eqnarray}
\frac{d\sigma}{dk_{T1}dy_1dz}&=&2\pi
k_{T1}\frac{1}{\pi S}\sum_{i,j,l}\int^1_{1-V+V W}\frac{dz'}{z'^2}
\int^V_{VW}\frac{dv}{1-v}f^A_i(x_1,M^2)f^B_{j}(x_2,M^2)
\frac{d\hat{\sigma}^{ij\rightarrow\gamma_2 l}}{dv}\nonumber \\
& &D_{\gamma_1/l}(z',M_F^2)\theta(z-1)\delta(\frac{1}{z}-z'),
\end{eqnarray}
and 
\begin{eqnarray}
\frac{d\sigma}{dk_{T1}dy_1dz}&=&2\pi
k_{T1}\frac{1}{\pi S}\sum_{i,j,k}\int^1_{1-V+V W}\frac{dz'}{z'^2}
\int^V_{VW}\frac{dv}{1-v}f^A_i(x_1,M^2)f^B_{j}(x_2,M^2)
\frac{d\hat{\sigma}^{ij\rightarrow\gamma_1 k}}{dv}\nonumber \\
& &D_{\gamma_2/k}(z',M_F^2)\theta(1-z)\delta(z-z'),
\end{eqnarray}
where $i,j$ and $k$ run over the various quark flavours and the gluon
and $M_F^2$ is the fragmentation scale. 

In the polarized case the cross sections are given by the same
expressions as above, but now the hard subprocess cross sections must be
replaced by the corresponding polarized ones, $d\Delta\hat{\sigma}/dv$.
The parton distributions must also be replaced by the corresponding
polarized ones which are defined by
\begin{equation}
\Delta f^A_a(x,Q^2)=f^A_{a,+}(x,Q^2)-f^A_{a,-}(x,Q^2)
\end{equation}
where $f^A_{a,\pm}(x,Q^2)$ is the distribution of parton of type $a$ with
positive $(+)$ or negative $(-)$ helicity in hadron $A$. Likewise the
polarized subprocess cross sections are defined by
\begin{equation}
\frac{d\Delta\hat{\sigma}}{dv}=\frac{d\hat{\sigma}}{dv}(+,+)-
\frac{d\hat{\sigma}}{dv}(+,-).
\end{equation}
Again, $+,-$ denote the helicities of the incoming partons. Since only
the initial state is polarized, we do not need to change the photon
fragmentation functions. The hard subprocess cross sections in LO
for these can be found, for example, in Ref.\cite{nlo2}. 

\subsection{The Box Diagram}

At collider energies, the box diagram process,
$gg\rightarrow\gamma\gamma$ (Fig.1c), although it is
$O(\alpha_s^2\alpha_{em}^2)$, has been shown to contribute significantly
to the cross section for unpolarized double prompt photon production in the 
low $k_T$
region due to the large gluon distribution at low-$x$. 
It is questionable whether or not one should include contributions from
the box diagram in an $O(\alpha_s)$ NLO calculation, since it is of
higher order, and may introduce numerical cancellations with the genuine
NLO contributions leading to misleading results. In a proper
phenomenological study at NLO we would not include this contribution,
but in order to make contact with the work of Ref.[12], and to estimate the
contribution from this process, we will include it.
The unpolarized
hard subprocess has been used many times before and can be found, for
example, in Ref.\cite{owens}. For the polarized case the hard subprocess
cross section can be obtained from the amplitudes according to the
prescription given in Ref.\cite{rd}.
To calculate the contribution form this process, we use a
similar expression to Eq.(2.5), with the quark distributions replaced by
gluon distributions and the hard subprocess cross section replace by
that for the $gg\rightarrow \gamma\gamma$ one. 

\subsection{The NLO contributions}

At NLO $O(\alpha_{em}\alpha_s)$ there are the virtual gluon corrections
to the LO $q\bar{q}$ annihilation process, plus real gluon emission
corrections $q\bar{q}\rightarrow\gamma\gamma g$ (Fig.(2)). In addition,
the 3-body process $qg\rightarrow\gamma\gamma q$ also
contributes (Fig(3)). The virtual contributions can be obtained directly
from Ref.\cite{nlo2}, where they were calculated for single prompt photon
production in both the polarized and unpolarized cases, by simply removing 
the non-abelian couplings. The 3-body matrix elements can be obtained in a 
similar way. 

A note concerning the matrix elements for the polarized case is in order
here. When taking the traces to calculate the helicity dependent matrix 
elements, one needs to project onto definite helicity states, labelled
$h$ for quarks and $\lambda$ for gluons, for the incoming particles.
This is achieved for quarks by defining spinors of definite helicity
according to
\begin{equation}
u(p_1,h)\bar{u}(p_1,h)=\frac{1}{2}\gamma^{\mu}(p_1)_{\mu}(1-h\gamma_5)
\end{equation}
where $p$ is the momentum. For gluons the general expression is
\begin{equation}
\epsilon_{\mu}(p_2,\lambda)\epsilon_{\nu}^*(p_2,\lambda)=\frac{1}{2}\left[
-g_{\mu\nu}+i\lambda\epsilon_{\mu\nu\rho\sigma}\frac{p_2^\rho p_1^\sigma}
{p_1.p_2}\right].
\end{equation}
This introduces the $\gamma_5$ matrix and the antisymmetric tensor
$\epsilon_{\mu\nu\rho\sigma}$ which are defined in $4$-dimensions. To work
in $n\neq4$-dimensions, some consistent scheme must be chosen in which
to treat
these $4$-dimensional objects. In Ref.\cite{nlo2}, the original scheme of
t'Hooft and Veltman \cite{thv}, and Breitenlohner and Maison \cite{bm} 
was chosen.
Briefly, this scheme makes a division of $n$-dimensional Minkowski space
into a $4$-dimensional and a $(n-4)$-dimensional part. Thus an arbitrary
vector $p$ will have a $4$-dimensional part, $\hat{\hat{p}}$, and an
$(n-4)$-dimensional part, $\hat{p}$. This means that scalar products of
these `hat' momenta will inevitably be present when traces are taken.
These terms in the matrix elements must be treated properly when phase
space integrals are performed. We shall consider this point in detail in
appendices B and C.   

In order to obtain a cross section differential in $k_{T1}$, $y_1$ and
$z$, we need to integrate 3-body phase space over the kinematic
variables of the unobserved quark or gluon. The method of performing
these integrals have been detailed in Ref.\cite{ael}. We outline the method
again in Appendix B and extend it to the polarized case where `hat'
momenta must also be considered. As in Ref.\cite{ael}, we restrict 
ourselves to
the cases where the photons are in opposite hemispheres, i.e. $z\geq 0$.
The phase space integrals are performed in $4-2\epsilon$ dimensions in
order to expose soft and collinear singularities as poles in $\epsilon$. 

In the case of $q\bar{q}$ scattering, we obtain single and double poles
in $\epsilon$ from the 3-body  phase space integrals. When the virtual
contributions are added to this the double poles automatically cancel
and the remaining single poles must be absorbed into the parton
distribution functions. Thus there are three pieces to our calculation
for this process which must be combined in order to get a finite
partonic subprocess cross section. There is the virtual contribution
which we represent by the function
\begin{displaymath}
\frac{d\sigma^V_{q\bar{q}} }{dvdwdz}\left(\hat{s},v,z,\frac{1}{\epsilon^2},
\frac{1}{\epsilon}\right).
\end{displaymath}
There is the result of integrating the 3-body matrix element over the
phase space of the unobserved partons as described in appendices B and
C, which we denote by the function
\begin{displaymath}
k'_{q\bar{q}}\left(\hat{s},v,w,z,\frac{1}{\epsilon^2},
\frac{1}{\epsilon}\right).
\end{displaymath}
and there is the factorization counter term 
\begin{eqnarray}
\frac{1}{\hat{s}v}\frac{d\sigma^F_{q\bar{q}}}{dvdwdz}&=&-
\frac{\alpha_s}{2\pi}\left[
\frac{1}{1-v w}H_{qq}\left(\frac{1-v}{1-v w},M^2\right)
\frac{d\sigma^{q\bar{q}\rightarrow\gamma\gamma}}{dv}\left(\frac{1-v}{1-v
w}\hat{s},v w,\epsilon\right)\delta(1-z)\right. \nonumber \\
& &\left. \frac{1}{v}H_{qq}(w,M^2)
\frac{d\sigma^{q\bar{q}\rightarrow\gamma\gamma}}{dv}
(w\hat{s},v,\epsilon)\delta(1-z)\right],
\end{eqnarray}
where
\begin{equation}
H_{ij}(z,Q^2)=-\frac{1}{\hat{\epsilon}}P_{ij}(z)\left(\frac{\mu^2}{Q^2}\right)+
f_{ij}(z).
\end{equation}
In the $\overline{MS}$ factorization scheme which we adopt,
$1/\hat{\epsilon}=1/\epsilon-\gamma_E+\ln 4\pi$, $f_{ij}(z)=0$, and
$P_{ij}(z)$ are the well known one-loop splitting functions for parton
$j$ into parton $i$ \cite{alt}. The finite subprocess cross section is now
obtained by adding these three parts,
\begin{eqnarray}
K_{q\bar{q}}(\hat{s},v,w,z,M^2)&=&k'_{q\bar{q}}\left(\hat{s},v,w,z,\frac{1}
{\epsilon^2},\frac{1}{\epsilon}\right)+\frac{d\sigma^V_{q\bar{q}}}
{dvdwdz}\left(\hat{s},
v,z,\frac{1}{\epsilon^2},\frac{1}{\epsilon}\right) \nonumber \\
&+&\frac{1}{\hat{s}v}\frac{d\sigma^F_{q\bar{q}}}{dvdwdz}
\left(\hat{s},v,w,z,\frac{1}{
\epsilon},M^2\right),
\end{eqnarray}
where $M^2$ is the factorization scale.

A similar procedure is followed for the $qg$ initiated process, except
now there are now no virtual contributions and hence no double poles.
Also different singularities are encountered when integrating the 3-body
matrix elements, hence we have a different factorization formula;
\begin{eqnarray}
\frac{1}{\hat{s}v}\frac{d\sigma^F_{qg}}{dvdwdz}&=&-\frac{\alpha_s}{2\pi}\left[
\frac{1}{1-v+v w}\tilde{H}_{\gamma q}\left(1-v+v w,M_F^2\right)
\frac{d\sigma^{gq\rightarrow\gamma q}}{dv}\left(
\hat{s},\frac{1-v}{1-v+v w},\epsilon\right)\delta(z_1-z)\right. \nonumber \\
& &\left. \frac{1}{v}H_{qg}(w,M^2)\frac{d\sigma^{q\bar{q}
\rightarrow\gamma\gamma}}{dv}(w\hat{s},v,\epsilon)\delta(1-z)\right.
\nonumber \\
&+&\left. \frac{1}{v}\tilde{H}_{\gamma q}\left(z,M_F^2\right)
\frac{d\sigma^{gq\rightarrow\gamma q}}{dv}\left(
\hat{s},v,\epsilon\right)\delta(1-w)\theta(1-z)\right].
\end{eqnarray}
where 
\begin{equation}
\tilde{H}_{\gamma q}(x,M_F^2)=-\frac{1}{\hat{\epsilon}}P_{\gamma q}(z)\left(
\frac{\mu^2}{M_F^2}\right),
\end{equation}
$M_F$ is the fragmentation scale and $z_1=1/(1-v+v w)$. 

Once $(\Delta)K_{q\bar{q}}$ and $(\Delta)K_{qg}$ have been calculated then we 
can use them to calculate the physical cross sections by
convoluting them with the appropriate parton distribution functions. In both the
polarized and unpolarized cases the general formula is 
\begin{eqnarray}
\frac{d(\Delta)\sigma}{dk_{T1}dy_1dz}&=&2\pi k_{T1}\frac{1}{\pi
S}\sum_{i,j}\int^1_{VW}\frac{dv}{1-v}\int^1_{VW/v}\frac{dw}{w}f^A_i(x_1,M^2)
f^B_j(x_2,M^2)\nonumber \\
& &\left[\frac{1}{v}\frac{d(\Delta)\hat{\sigma}^{ij}}{dv}\delta(1-z)\delta(1-w)+
\frac{\alpha_s(\mu^2)}{2\pi}(\Delta)K_{ij}(\hat{s},v,w,z,M^2,M_F^2)\right].
\end{eqnarray}
The indices $i,j$ run over quark flavours or represent a gluon, and in
the case of $q\bar{q}$ scattering, the first term is the LO
contribution, which is absent for the $qg$ initiated process.

\section{Monte Carlo Calculation}

The combination of analytic and Monte Carlo techniques used here to
perform the phase space integrals has been documented and described in
detail elsewhere (see Bailey {\it et al.,}\cite{ael} and references therein), 
so our discussion will be fairly brief,
highlighting mostly those features which are important to our
calculation. The basic technique comprises isolating those
regions of phase space where soft and collinear singularities occur and
integrating over them analytically in $4-2\epsilon$ dimensions. In this
way the singularities are exposed as poles in $\epsilon$. These regions
are isolated form the rest of the 3-body phase space by the imposition
of arbitrary boundaries between them, achieved by introducing cut-off
parameters. The soft gluon region of phase space is defined to be the
region where the gluon energy, in a specified reference frame, usually
the subprocess rest frame, falls below a certain threshold,
$\delta_s\sqrt{\hat{s}}/2$, where $\delta_s$ is the arbitrary cut-off
parameter and $\hat{s}$ is the center-of-mass energy in the
parton-parton system. If we label the momenta for the general 3-body
process by $p_1+p_2\rightarrow p_3+p_4+p_5$, then we can define the
general invariants by $s_{ij}=(p_i+p_j)^2$ and $t_{ij}=(p_i-p_j)^2$. The
collinear region is then defined as the region where the value of an invariant
falls below the value $\delta_c \hat{s}$. 

The phase space integration over the mutually exclusive soft and
collinear regions are performed not on the full matrix elements but on
approximate versions of them defined in specific ways. In the soft gluon case
they are obtained by setting the gluon energy to zero everywhere where
it occurs in the matrix elements, except in the denominators. This is the soft 
gluon approximation. Similarly, in the collinear case, each 
invariant which vanishes is in turn set to zero everywhere except in the
denominator. This is the leading pole approximation. The phase space
integrals are then performed on these and only the logarithms of the
cut-off parameters are retained. All positive powers of the cut-off
parameters are set to zero. The meaning of all this is that for the method to 
work the parameters must be kept small, otherwise the approximations made
will no longer be valid and the method would fail. 

Once the phase space integrals have been performed and the soft and
collinear poles are exposed, then the virtual contributions, if any, are
added to them, at which time all double poles and single poles of soft
(IR) origin automatically cancel. The remaining collinear poles are then
factorized in the parton distribution and fragmentation functions in the
usual way, at some scale and using some specific factorization scheme.
In our case we use the $\overline{MS}$ scheme. At this point 
one is left with a set of 2-body processes which depend explicitly on
$\ln\delta_s$ and $\ln \delta_c$, and a set of 3-body matrix elements
which when integrated over the phase space using Monte Carlo techniques,
have an implicit dependence on these same logarithms, but which now have 
opposite signs in order to cancel the dependence in the 2-body part. The 
physical cross sections will then be independent of these arbitrary cut-off 
parameters.

It is generally very simple to impose experimental cuts and to calculate
different observables when one is dealing with 2-body matrix elements
only, as in the usual LO calculations. This is not the case when we
consider 3-body processes, as there is the need now to cancel soft and
collinear divergences, but beyond that, the standard techniques require
often complex Jacobian transformations to calculate cross sections
differential in different variables, and the phase space integrals can 
sometimes only be done analytically when specific limits of integration
are involved. This can sometimes make in impossible to impose cuts on
the kinematic variables. Thus, fully analytic methods of performing
calculations for physical processes, although in some cases desirable,
can be rather restricted in their usefulness when confronted with
experimental situations where it is often desirable and sometimes
even unavoidable that cuts be made. The combined analytic and Monte
Carlo method does not suffer from these particular drawbacks, in that it
is very easy to calculate cross sections differential in many different
variables at once, and cuts on the kinematic variables to match those
made in the experiments can be easily imposed because the phase space 
integrals are performed numerically after all singularities have been 
dealt with.

Our calculation of double prompt photon production proceeds along 
exactly the same lines to that described in the second of Ref.\cite{ael}. 
In fact the only difference is that in the polarized case we must deal
with 'hat' momentum integrals as outlined in section II. It turns out
that the 'hat' momenta only contribute in the collinear limit when an
initial state parton splits into two partons, one of which participates
in the hard scattering. This point is discussed in fair detail in
Ref.\cite{nlo2}, where it was suggested that, since these contributions 
seem to
be of the same form for a particular vertex in the collinear limit, they
may be regarded as universal properties of these parton legs in the HVBM
scheme, and as such may be factorized into the parton distributions
under a new ($\overline{MS}_p$) factorization scheme. In our
calculation, we choose not to use this scheme because the new polarized
parton distributions are being calculated in the standard $\overline{MS}$
scheme. Thus, these contributions are included in our 2-body part on the
cross section.  

Thus all the discussion in Ref.\cite{ael} on the Monte Carlo method can be
carried directly over to the polarized case, expect that once
factorization has been performed the expression for the remnants of the
hard collinear singularities is now given for the polarized case by
\begin{eqnarray}
\frac{d\Delta\tilde{\sigma}}{dv}(q\bar{q}\rightarrow\gamma\gamma)=
\frac{\alpha_s}{2\pi}\frac{d\Delta\sigma^{Born}}{dv} & & \nonumber \\
&\times&\left[ \Delta f^A_q(x_1,M^2)\int^{1-\delta_s}_{x_2}\frac{dz}{z}
f^B_{\bar{q}}(x_2/z,M^2)\Delta\tilde{P}_{qq}(z)\right. \nonumber \\
&+&\left. \Delta f^A_q(x_1,M^2)\int^{1}_{x_2}\frac{dz}{z}
f^B_{g}(x_2/z,M^2)\Delta\tilde{P}_{qg}(z)\right. \nonumber \\
&+&\left. \Delta f^B_{\bar{q}}(x_2,M^2)\int^{1-\delta_s}_{x_1}\frac{dz}{z}
f^A_{q}(x_1/z,M^2)\Delta\tilde{P}_{qq}(z)\right. \nonumber \\
&+&\left. \Delta f^B_{\bar{q}}(x_2,M^2)\int^{1}_{x_1}\frac{dz}{z}
f^A_{g}(x_1/z,M^2)\Delta\tilde{P}_{qg}(z) \right]
\end{eqnarray}
with
\begin{equation}
\Delta\tilde{P}_{ij}(z)=\Delta
P_{ij}(z)\ln\left(\frac{1-z}{z}\delta_c\frac{\hat{s}}{M^2}\right)-\Delta
P_{ij}'.
\end{equation}
The polarized Altarelli-Parisi splitting functions in $4-2\epsilon$
dimensions are
\begin{equation} 
\Delta P_{qq}(z,\epsilon)=C_F\left[\frac{1+z^2}{1-z}+3\epsilon
(1-z)\right],
\end{equation}
and
\begin{equation}
\Delta P_{qg}(z,\epsilon)=\frac{1}{2}\left[
(2z-1)-2 \epsilon (1-z)\right].
\end{equation}
Part of the second terms in the splitting functions come from the 'hat'-momenta
as discussed in Ref.\cite{nlo2}. The functions $\Delta P'_{ij}$ are
defined by the relation,
\begin{equation}
\Delta P_{ij}(z,\epsilon)=\Delta P_{ij}(z)+\epsilon \Delta P'_{ij}(z).
\end{equation}

\section{Numerical Results}

In this section we present some  numerical results for polarized and
unpolarized double prompt photon production at RHIC center-of-mass
energies. We first study the non-isolated cross section, since we do not
know what isolation restrictions will be necessary at RHIC. We will then
present some results for the isolated cross section, assuming some
plausible isolation parameters, based on those used by the CDF 
collaboration at Fermilab. 

Throughout we use the GRV \cite{grv} parton
distributions for the proton in the unpolarized case. We use
$\Lambda^{(4)}_{QCD}=0.200$, to match the GRV parton distributions, but
we do not include any contribution for charm quarks, since these are not 
included in the polarized parton distributions. For the polarized parton
distributions we use the two sets proposed by Cheng and Wai
\cite{chengwai}. In the first set which we will refer to as scenario
$a$, a large polarized gluon distribution is assumed, and the SU(3)
flavour symmetric sea quark distribution is assumed to vanish at the
input scale. In the other case (scenario $b$), the gluon distribution 
vanishes at the
input scale, but the SU(3) flavour symmetric polarized sea quark
distribution is assumed to be directly related to the unpolarized
strange sea, leading to a large negatively polarized sea, and small
polarized gluon distributions. In both cases the valence distributions
are assumed proportional to the unpolarized ones. The main aim of this
paper is not to provide the most up to date phenomenological study of
the polarized cross section, but to gauge the importance of higher order 
corrections and make contact with the results of Ref.[12]. We are also
interested in whether the cross section is sensitive to $\Delta G$,
hence the use of the Cheng and Wai distributions which have extreme
gluon distributions.

For the electromagnetic coupling constant we use $\alpha_{em}=1/137$ and
we use the two loop expression for $\alpha_s(\mu^2)$. Unless otherwise 
stated, we set all factorization/renormalization scales to
$\mu^2=((k^{\gamma}_{T1})^2+(k^{\gamma}_{T2})^2)/2$. We make use of the
asymptotic parametrizations for the parton to photon fragmentation
functions provided in Ref.\cite{owens}. We assume the maximum
$\sqrt{S}=500\;GeV$ for the RHIC centre-of-mass energy.

In a future study \cite{cg} we will use photon fragmentation functions
evolved in NLO, polarized parton distributions evolved in NLO as well as
including a contribution from charm quarks. There we hope to make
up-to-date estimates of the cross section ate RHIC.
 
\subsection{The Inclusive Cross Section}

In the analytic calculation of the inclusive cross section, we exposed
the various soft and collinear poles by expanding the integrated matrix
elements as plus-distributions in the variable $z$. The details are
given in the Appendix. This procedure ensures that these integrable
singularities can be treated numerically, but it prevents us from
providing distributions in the $z$-variable with infinitely sharp
resolution. That is, in order for us to present finite $z$-distributions
in we must integrate over some finite range of $z$. Following the
procedure used in Ref.\cite{ael}, we define
\begin{equation}
\frac{d(\Delta)\sigma}{dk^{\gamma}_{T1}dy_1dz}=\frac{1}{\Delta
z}\int^{z+\frac{\Delta z}{2}}_{z-\frac{\Delta z}{2}}  
\frac{d(\Delta)\sigma}{dk^{\gamma}_{T1}dy_1dz'} dz',
\end{equation}
and provide the distributions in $z$ with finite bin widths $\Delta z$.
For the $k_T$ distributions we integrate over a specified range of $z$,
\begin{equation}
\frac{d(\Delta)\sigma}{dk^{\gamma}_{T1}dy_1}=
\int^{z_b}_{z_a}  
\frac{d(\Delta)\sigma}{dk^{\gamma}_{T1}dy_1dz} dz.
\end{equation}

In fig.4a we show the $z$-distribution for the unpolarized cross
section, at $k_{T1}^{\gamma}=5\;GeV$ and $y_1=0$, for a bin size
$\Delta z=0.2$. The LO Born term gives a sharp peak at $z=1$ but the
effect of the higher order corrections is to reduce and broaden this
peak. The higher order correction terms, neglecting the $gg$ process
which is positive and contributes only at $z=1$, give a negative contribution 
at $z=1$, as indicated by the dashed line. 
It turns out that the fragmentation terms where the trigger photon is
produced via fragmentation give large contributions in the region 
$z\leq 1$, whereas those where the other photon is produced in this way
give a positive but not as large contribution at $z\geq 1$. This is
partly due to the fact that phase space runs out in this region.
The overall shape of the distribution is similar to that
obtained in Ref.\cite{ael}. 

In Fig.4b we compare the polarized and unpolarized cross sections with
the same parameters as in fig.4a. The unpolarized cross section is
scaled by a factor of $1/10$ for easier comparison. In general the
curves have a similar shape, but due to cancellation between various
parts of the polarized cross section, it has a much more sharply
pronounced peak at $z=1$. 

The $k_T$ distribution is displayed in fig.5a for $0.2\leq z \leq 2.0$
and $|y_1|\leq 3$. The solid curve is the unpolarized cross section,
while the dashed and dotted curves are the polarized cross sections for
scenario $a$ and $b$ respectively. Clearly, given a reasonably large
luminosity, the cross sections are large enough to measure out to a
$k_T$ of about $50\;GeV$ say. Also the two scenarios give significantly 
different predictions indicating sensitivity to the polarized
distributions. The main question is the sensitivity of
the cross section to polarization. To  determine this we plot the
longitudinal asymmetry $A_{LL}$ defined by
\begin{equation}
A_{LL}=\frac{ \frac{d\Delta \sigma}{dk_{T1}^{\gamma}dy^{\gamma}_1}}  
{ \frac{d\sigma}{dk_{T1}^{\gamma}dy^{\gamma}_1}}.
\end{equation}
The dotted curve shows the asymmetry predicted for the Born cross
section using the parton distributions of scenario $a$. At the hard
parton-parton scattering level this asymmetry is $A_{LL}=-1$, and is
thus modified significantly by folding with the parton distributions.
Note that this is not the asymmetry for the LO cross section, since we
have not included the contributions from photon fragmentation. 
The solid and dashed curves give the asymmetry of the full higher cross
section for scenario $a$ and $b$ respectively. In the case of scenario
$a$, the asymmetry is already about $10\%$ at $k_T=15\;GeV$, and rises to
nearly $20\%$ at $k_T=50\;GeV$. For scenario $b$ the asymmetry is more
modest, varying between $4$ and $9\%$ for a similar range in $k_T$.
Thus, the inclusive cross section is sensitive to the polarized parton
distributions used, and the asymmetry not too small in some accessible
kinematic regions. 

Of course as we have stated before the inclusive cross section we have
calculated is probably not going to be of much practical use at the
RHIC collider due to the need to isolate in order to identify the photon
signal. Its main usefulness is as a check on the more versatile
calculation using Monte Carlo techniques which we shall discuss in the
next section. 

\subsection{The Isolated Cross Section} 

In this section we study the isolated cross section for double prompt photon
production at RHIC. We keep all distributions and parameters the same as
for the inclusive case, but in addition a cone size of $R=0.7$ and an
energy resolution parameter $\epsilon=2\;GeV/k_T^{\gamma}$ is used. 
The cone of radius $R$, with the photon at the centre, is defined in the 
pseudorapidity-azimuthal angle plane ($y-\phi$-plane) by the relation
\begin{displaymath}
R=\sqrt{(\Delta y)^2+(\Delta \phi)^2}.
\end{displaymath}
If hadronic energy greater than a fraction $\epsilon$ of the photon
energy is observed in the cone then the event is rejected. This serves
to define the isolated cross section. 

In fig. 6a we display the $k_T$ distribution of the full isolated polarized
and unpolarized cross sections. Note that this curve cannot be directly
compared with that in fig.5a since here the cuts are different. In
fig.6a both photons are still allowed to have rapidities in the range
$|y^{\gamma}|\leq 3$, but for a fixed $k_T$ of one of the photons which
we plot vs cross section, the other is allowed to have $k_T\geq 5\;GeV$.
Our predictions indicate a measurable cross section out to $k_T$ around
$30$ to $40\;GeV$, given enough luminosity. Again, in the polarized case, 
sensitivity to the
polarized parton distribution chosen is evident from the figure. 

Fig.6b shows the asymmetries of the cross sections in fig.6a for scenarios 
$a$ and $b$. There is a very pronounced difference in the asymmetries
for scenarios $a$ vs $b$, indicating a corresponding sensitivity to the
polarized parton distributions.  The implication of these results is
that discrimination between the two extreme cases presented here should
be possible from measurement of this cross section with moderately good 
statistics at RHIC. In fig.6c we break down the asymmetry in the various 
initial state contributing processes for the case of scenario $a$. The $gg$ 
initiated process, as expected, only gives a significant contribution at
low $k_T$ due to the steeply falling gluon distribution. But we find
that for this scenario the asymmetry is dominated by the $qg$ initiated
process. We did a similar analysis for scenario $b$  and found that the
contribution from the $q\bar{q}$ initiated process was largest and that
from the $gg$ initiated process was completely negligible. The cross
section is thus very sensitive to the size of $\Delta G$.

In fig.7 we show the scale dependence of the isolated cross section,
with the same cuts as above, for
the unpolarized case. We chose three different scales
$\mu^2=n^2( (k_{T1}^{\gamma})^2+(k_{T2}^{\gamma})^2)/2$, where $n=1,1/2$
and $2$. All
factorization/renormalization scales are varied simultaneously. There is
obviously some dependence of the cross section on the scales chosen,
leading to an estimated $20\%$ uncertainty in our predictions at
$k_T=15\;GeV$.

Lastly in fig.8 we show the $k_T$ distribution of the cross section for 
unpolarized isolated double prompt photon production for the same cuts chosen
above but at a cms energy of $\sqrt{S}=14\;TeV$ appropriate for the LHC
collider. The cross section is as expected much larger than at lower
energies and even at $k_T=100\;GeV$ we can expect a cross section of
more than $0.1\;pb$. 

\section{Conclusions}

In this paper we calculated the cross section for polarized and unpolarized
double prompt photon production in $pp$ collisions at cms energies
appropriate for the RHIC collider. We examined whether the isolated
cross section is sensitive to the spin dependent gluon distributions of
the proton, by comparing two extreme (LO) parametrizations of $\Delta G$. Our 
results indicate that the cross section is large enough
to measure, and that it should indeed be sensitive to the polarized gluon
distributions leading to the possibility that it may be useful to help
discriminate between some extreme scenarios for the polarized
distributions. Our results differ from those reached in a previous
,mostly LO, study in that we find a significant contribution to the
asymmetry from the $qg$ initiated process, leading to a corresponding
sensitivity of the cross section to the polarized gluon distribution,
$\Delta G$. This cross section should therefore be useful as a
supplement for information on the polarized distributions gathered from
more sensitive sources such as jet or prompt photon production. The main 
drawback of our study is the use of rather outdated
polarized parton distributions evolved in LO only. We therefore do not
claim to have provided NLO estimates for the polarized cross section at
RHIC, but only an indication as to the size of the cross section and
sensitivity to $\Delta G$. 

Indications are that the cross section for double prompt
photon production will, as expected, be sizable at the LHC, and will
thus provide a significant background to Higgs searches.

\section{Acknowledgements}
C.C. thanks the Theory Group at Argonne and in particular Alan White 
for their kind hospitality during the final stage of this work. 
We thank E. L. Berger, G. Bodwin, R. Field, A. White, W. Vogelsang  and
G. Ramsey for 
comments and discussions on the matter presented, and W. Vogelsang for 
reading the manuscript. We thank P. Ramond for illuminating discussions 
concerning the role of the anomalies.
 Finally we express our warm gratitude to Derek and 
Liz Kruk, Deb Petrie  and Betsy Herman for encouragement.
This work supported in part by the U.S.~Department of Energy, Division 
of High Energy Physics, Contract W-31-109-ENG-38 and DEFG05-86-ER-40272.

\pagebreak

\appendix
\section{Regularization schemes}

We use the $\gamma_5$ scheme of Breitenlohner and Maison \cite{bm}, which is 
free of internal inconsistencies in the definition of $\gamma_5$ 
in $D-$dimensions. 
An anticommuting $\gamma_5$ is not compatible with Dimensional Regularization
\cite{thv,bm,bonn}. In the t'Hooft-Veltman scheme, systematized by 
Breitenlohner and Maison, an $n$-dimensional $\gamma_\mu$ matrix is split 
into its 
4-dimensional component $\widehat{\widehat{\gamma}}_\mu$ and a remaining 
component $\widehat{\gamma}$. Thus 
$\gamma_\mu=\widehat{\widehat{\gamma}}_\mu \,\,+\,\, \widehat{\gamma}_\mu$.
A suitable representation of $\gamma_5$ is 
\beq
\gamma_5={i\over 4}\epsilon_{\alpha\beta\gamma\delta}
\gamma^\alpha\gamma^\beta\gamma^\gamma\gamma^\delta. 
\eeq

If it is postulated that $\gamma_5$ anticommute with the other 4-dimensional 
Dirac matrices 
and anticommute with the remaining ones, then
all the corresponding algebraic relations can be shown to be 
consistent with dimensional regularization \cite{bm}.
Therefore we  {\em define}
\beqa
&& \gamma_5\widehat{\widehat{\gamma}}_\mu + \widehat{\widehat{\gamma}}_\mu 
\gamma_5=0
\nonumber \\
&& \gamma_5\widehat{\gamma}_\mu - \widehat{\gamma}_\mu\gamma_5=0.
\nonumber \\
\eeqa
A discussion of the modifications which appear in the integration over the phase 
space of the 3 final states, due to this ansatz, is presented in the 
next section.

\section{3-particle Phase Space for Polarized Scatterings}
As we have discussed before, the use of the t'Hooft-Veltman regularization 
\cite{thv} introduces a dependence of the matrix elements on the hat-momenta 
which requires, in part, a modification of the phase space integral which appear in the unpolarized case. 

In the case of unpolarized scattering,
 the singularities are generated by poles 
in the matrix elements which have the form $1/t_3$, $1/u_3$, $1/(t_3 u_3)$ and 
similar ones, in multiple combinations of them. Multiple poles can be reduced 
to sums of combinations of double poles by using simple identities  
among all the invariants and by the repeated use of partial fractioning. 
This is by now a well established procedure. 
In our case we encounter new terms of the 
form $1/t_3^2$ and $1/u_3^2$ and new matrix elements containing typical factors 
of the form ${\widehat{\widehat{k}}}_3$,  
$\widehat{\widehat{k}}_2$, and $\widehat{\widehat{k_2\cdot k_3}}$
at the numerator. Let's discuss for a moment these last terms containing 
hat-momenta. It is obvious that by a suitable choice of the parametrizations 
given by the sets 1, 2 ,3 and 4, (defined in the next section) we are able 
to reduce to the ordinary phase 
space result given by (\ref{ps3}) all the matrix elements containing scalar 
products of the form $\widehat{\widehat{p_i\cdot k_j}}$, 
$\widehat{\widehat{k_1\cdot k_j}}$ 
with $i=1,2$, $j\,=1,\,2,\,3$. Therefore it is possible to set to zero, 
after taking the traces, all the products contain such combinations of
 hat-momenta. Then, the only 
matrix elements of hat-momenta which are left and which are not 
set to zero are those containing products of the form 
$\widehat{\widehat{k_a\cdot k_b}}$,with $a,\,b=\,1,\,2$.

Let's consider the 3-particle phase space integral when hat-momenta are 
present

\beqa
PS_3\equiv \int d^n k_1 d^n k_3 d^n k_2 \delta(k_3^2)\delta(k_2^2)
\delta(k_1^2)\delta^n(p_1 +p_2 - k_3 -k_2 -k_1)\widehat{\widehat{k_2}}^2
\label{ps1}
\eeqa
and let's lump together the momenta $k_3$ and $k_2$ as follows

\beqa
&& PS_3=\int d^n k_1 d^n k_3 d^n k_2 d^n k_{23} \delta(k_3^2)\delta(k_2^2)
\delta(k_1^2)\delta^n(p_1 +p_2  -k_2- k_1)\nonumber \\
&& \delta( p_1 + p_2 -k_1 - k_{23})
\delta^n(k_3 + k_2 - k_{23})\widehat{\widehat{k_2}}^2.
\label{ps2}
\eeqa

Eq.~(\ref{ps2}), once integrated over $k_{23}$, gives (\ref{ps1}). 
We single out the invariant mass of the pair $(1,2)$ by the relation 
\beq
1=\int dm^2 \delta(m^2- k_{23}^2)
\eeq
 which we insert in (\ref{ps2}) to get
\beqa
&&\int d^n k_1 d^n k_3 d^n k_2 d^n k_{23} d m^2 \delta(k_3^2)\delta(k_2^2)
\delta(k_1^2) 
\delta^n( p_1 + p_2 -k_1 - k_{23})\nonumber \\
&&\delta^n(k_3 + k_2 - k_{23})\delta(m^2- k_{23}^2)\widehat{\widehat{k_2}}^2.
\eeqa

Notice that by this trick we can factorize a 2-particle phase space 

\beq
PS_3=\int d^n k_1 d m^2 d^n k_{23}\delta(k_1^2)\delta(k_{23}^2-m^2)
\delta^n(p_1 + p_2 - k_1 - k_{23})\times PS_2
\label{trick}
\eeq

\beq
PS_2\equiv \int d^n k_2\delta(k_2^2)\delta((k_{23}- k_2)^2)
\widehat{\widehat{k}}_2^2
\label{trickp2}.
\eeq
We have set $k_2=(\widehat{\widehat{k}}_2,\widehat{k}_2)$, with
\beq
\widehat{\widehat{k}}_2=k_2^0(1,\cos\theta_3 \sin\theta_2 \sin\theta_1,
\cos\theta_2 \sin\,\theta_1,\cos\,\theta_1)
\eeq

being the 4-dimensional part of $k_2$. 
We easily get 
\beq
\widehat{\widehat{k}}^2_2= {s_{34}\over 4}
\sin^2\theta_3\,\sin^2\theta_1\,\sin^2\theta_2.
\eeq

Therefore the usual angular integration measure 
\beq
d\Omega^{(n-2)}=\prod_{l=1}^{n-l-2}=
\sin^{n-l-2}\theta_l\,d\theta_l 
\eeq
 is effectively modified to 
\beq
d\Omega^{(n-2)}=\prod_{l=1}^{3}\sin^{n-l}\theta_1 d\theta_l\times 
\prod^{n-2}_{l=4} \sin^{n-l-2}\theta_l d\theta_l
\eeq

This integral is evaluated in a special frame. Assuming that 
$k_{23}^2>0$, we sit in the center of mass frame of the $(1,2)$ pair, 
in which  $k_{23}=(m, {\bf 0})$ and $k_2=(E_2, {\bf k_2})$ to get
\beqa
&& PS_2=\int d^n k_2 \delta(k_2^2)\delta(m^2-2 m E_2)\nonumber \\
&&=k_{23}^{n/2-1}I[\theta_i]
\label{trick3}
\eeqa
where
\beq
I[\theta_i]={\pi^{n/2-2}\over  \,2^{n}\Gamma[n/2-2]}
\int_{0}^{\pi}\sin^{n-1}\theta_1\, d\theta_1\int_{0}^{\pi}d\theta_2 
\sin^{n-2}\theta_2\int_{0}^{\pi}d\theta_3 \sin^{n-3}\theta_3
\eeq

where $\theta_1$ and $\theta_2$ are the only relevant 
angles which appear in the matrix elements and therefore are not integrated. 
We have displayed also the $\theta_3$ integral since it is different from the 
unpolarized case.

After integration over $m^2$ and $k_{23}$ and a simple covariantization we get

\beqa
&& PS_3=
\int d^nk_1\delta(k_1^2)[( p_1 + p_2 -k_1)^2]^{n/2-1}\, I[\theta_i].
\eeqa
The collinear or infrared poles are therefore isolated in the c.m. 
system of the pair. The remaining part of 
the integral is evaluated in the c.m. system 
of the two incoming partons $p_1, p_2$.
At this point let's consider the sub-integral 
\beqa
&&I_1=\int d^n k_1\delta(k_1^2)[(p_1 + p_2-k_1)^2]^{n/2-1}\nonumber \\
&& ={\pi^{n/2-1}\over \Gamma[n/2-1]}
\int E_3^{n-3}d E_3 \sin^{n-3}\theta_1 d\theta_1 (s+t + u)^{n/2-1}.
\eeqa
We have chosen the parameterization 

\beq
k_1=E_3(1,..., \cos\theta_2\,\, \sin\theta_1,\cos\theta_1)
\label{k33}
\eeq

where the dots indicate $n-3$ components which we integrate over 
in a trivial way. 

We introduce the change of variables $(\cos\theta_1,E_3)\to (v,w)$ 
with $v$ and $w$ defined as in (\ref{vw}).
In the c.m system of the two incoming (massless) 
partons we have that $p_1= Q n^+$ and $p_2=Q n^-$
with $ Q=\sqrt{s/2}$
\beq 
n^{\pm}={1\over \sqrt{2}}(1,{\bf 0_\perp},\pm 1).
\eeq

From $t=(k_1-p_1)^2=-\sqrt{s}E_3(1-\cos\theta_1)$, 
$u=(k_1-p_2)^2=-\sqrt{s}E_3(1+\cos\theta_1)$ and using (\ref{vw})
we get 
  
\beqa
&& \cos\theta_1 ={v w - 1+v\over v w +1 -v}\nonumber \\
&& E_3={\sqrt{s}\over 2}(v w +1 -v) \nonumber \\
&& s+t +u=s v (1-w) \nonumber \\
&& (1-\cos^2\theta_1)={4 (1-v)v w\over (v w + 1-v)^2}
\eeqa
and the jacobian of the transformation to be
$\partial(\cos\theta_1, E_3)/\partial(v,w)=\sqrt{s}v/ (1-v + v w)$. 
Therefore we obtain 
 
\beqa
&&PS_3= {\pi^{n-5/2}\over  2^{n+1}\Gamma[n/2-1/2]\Gamma[n/2-2]}
s^{n-2}\int dv dw (1-v)^{n/2-2} 
(1-w)^{n/2-1}v^{n-2}w^{n/2-2}\nonumber \\ 
&&\,\,\,\,\,\,\,\,\,\,\,\,\,\,\,\,\,\,\,
\times \int_{0}^{\pi}d\theta_1 \int_{0}^{\pi}d\theta_2 \sin^{n-1}\theta_1
\sin^{n-2}\theta_2.
\eeqa
In order to compare this result with \cite{ellis} we need an
additional normalization factor $1/(2 \pi)^{5-4 \epsilon}$, due to the different 
definition of $PS_3$ we have adopted  (now $n=4-2\epsilon$)
and we get 
\beqa
&& Ps_3\equiv{1\over 2 \pi^{5-4 \epsilon}}PS_3\nonumber \\
&& ={s^{2-2\epsilon} \pi^{1-2\epsilon}\over 2^4 
(2 \pi)^{5-4\epsilon}}\left({-\epsilon\over (1-\epsilon)}\right)
2^{3-2 \epsilon}\int dv dw (1-v)^{-\epsilon} (1-w)^{1-\epsilon}v^{2-2 \epsilon}
w^{-\epsilon}\nonumber \\
&&\,\,\,\,\,\,\,\,\,\,\,\,\,\,\,\,\,\,\,\,
 \int_{0}^{\pi}d\theta_2 \sin^{2-2\epsilon}\theta_2\int_{0}^{\pi}d\theta_1 
\sin^{3-2\epsilon}\theta_1.
\eeqa

In order to integrate over the matrix elements, we need to evaluate the 
various scalar products which appear in such matrix elements,  in the 
c.m. frame of the pair $(1,2)$. For this purpose we define the 
functions 

\beqa
&& P[x,y,z]= ({x^2 + y^2 +z^2-2 x y -2 y z -2 x z\over 4 x})^{1/2}
\nonumber \\
&& E[x,y,z] ={x+y-z\over 2 \sqrt{x}}.
\eeqa
It is easy to show that 

\beqa
&& |p_1|=P[s_{23}, p_1^2,u_1]\nonumber \\
&& |p_2|=P[s_{23},p_2^2,t_1] \nonumber \\
&&|k_3|=|k_2|=P[s_{23},p_1^2,p_2^2]\nonumber \\
&& |k_1|=\sqrt{s\over s_{23}}P[s,k_1^3,s_{23}]\nonumber \\
&& p_1^0=E[s_{23},p_1^2,u_1]\nonumber \\
&&p_2^0=E[s_{23},p_2^2,t_1]
\label{list}
\eeqa
When all the external lines of the $2\to 3$ process are massless, then we 
specialize (\ref{list}) as follows 

\beqa
&& p_1^0=E[s_{23},0,u_3]={s_{23}-u_1\over 2 \sqrt{s_{23}}}\nonumber \\
&& p_2^0=E[s_{23},0, t_3]={s_{23}-t_3\over 2 \sqrt{s_{23}}}.\nonumber \\
\eeqa

Now, using $t_3=s(v-1)$ and $u_3=-s v w$
we get

\beqa
&& p_1^0={s v\over 2 \sqrt{s_{23}}}\nonumber \\
&& p_2^0={s(1-v w)\over 2 \sqrt{s_{23}}}
\eeqa

In the massless case, the energies of the external final state particles 
are given by 
\beqa
&& k_1^0=\sqrt{s\over s_{23}}P[s,0,s_{23}]\nonumber \\
&&={1\over \sqrt{ s_{23}}}
{(s-s_{23})\over 2}\nonumber \\
&& = {1\over 2 \sqrt{s_{23}}}s(1-v+w)\nonumber \\
&& k_3^0=k_2^0={\sqrt{s_{23}}\over 2}.
\label{k3}
\eeqa

In the derivation of (\ref{k3}) we have used the relation 
$s+ t_1 + u_1=s_{23}$ together with (\ref{vw}). 
There are four different parametrizations of the integration momenta 
which we will be using. In the first one, which is suitable for unpolarized 
scattering \cite{ael}, one defines (in the c.m. frame of the pair $(1,2)$)
\begin{itemize}
\item{set 1}
\beqa
&& k_1={1\over 2 \sqrt{s_{23}}}s (1-v+w)(1,0,...,\sin\,\psi,\cos\,\psi)
\nonumber \\
&& k_3={\sqrt{s_{23}}\over 2}(1,...,\cos\theta_2 \sin\theta_1,\cos\theta_1)
\nonumber \\
&& k_2={\sqrt{s_{23}}\over 2}(1,...,-\cos\theta_2 \sin\theta_1,-\cos\theta_1)
\nonumber \\
&& p_1={sv\over2 \sqrt{s_{23}}}(1,0,...,0,\sin\psi_2,\cos\psi_2)\nonumber \\
&& p_2={s(1-v w)\over 2 \sqrt{s_{23}}}(1,0,...0,\sin\psi_2,\cos\psi_2)
\label{kk}
\eeqa
\end{itemize}
where the dots denote the remaining $n-2$ polar components. 
Similarly, in the evaluation of the integrals over the hat-momenta we need the 
other parametrizations

\begin{itemize}
\item{set 2}
\beqa
&& p_1=p_1^0(1,0,...,0,0,1)\nonumber \\
&& p_2=p_2^0(1,0,...,-\sin\psi_2,0,\cos\psi_2)\nonumber \\
&& k_1=k_1^0(1,0,...,-\sin\psi_0,0,\cos\psi_0)\nonumber \\
\eeqa

\item{set 3}
\beqa
&& p_1=p_1^0(1,0,...,\sin\psi_2,0,\cos\psi_2)\nonumber \\
&& p_2=p_2^0(1,0,...,0,0,1)\nonumber \\
&& k_1=k_1^0(1,0,...,\sin\psi_1,0,\cos\psi_1)\nonumber \\
\eeqa

\item{set 4}
\beqa
&& p_1=p_1^0(1,0,...,\sin\psi_0,0,\cos\psi_0)\nonumber \\
&& p_2=p_2^0((1,0,...,-\sin\psi_1,0,\cos\psi_1)\nonumber \\
&& k_1=k_1^0(1,0,...,0,0,1)\nonumber \\
\eeqa
\end{itemize}
where $0,...$ refers to $n-5$ components identically zero. 
It is straightforward to obtain the relations 
\beqa
&& \sin\psi_0={2\sqrt{w (1-v)(1-w)}\over 1-v + v w}\nonumber \\
&& \sin\psi_1={2 v\sqrt{w(1-v)(1-w)}\over (1-vw)(1-v + v w)}
\nonumber \\
&& \sin\psi_2={2\sqrt{w(1-v)(1-w)}\over 1-v w}.
\eeqa
Which set of parametrizations we are going to use depends on the 
form of the hat-momenta which appear at the numerators of the matrix elements 
after the traces are performed. 

Notice that  
\beq
k_3+k_2=\sqrt{s_{23}}(1,{\bf 0}).
\eeq

In the c.m. of the pair $(1,2)$ 
$p_1$, $p_2$ and $k_1$ lie on a plane, with the spatial components 
satisfying the condition ${\bf p_1}+{\bf p_2}+{\bf k_1}=0$. 
Using (\ref{kk}) and the expressions of $t_1,u_1$ and $s_{23}=sv(1-w)$ 
in terms of $v$ and $w$, we can easily obtain the relations
\beqa
&& \sin\psi_{2}=\sqrt{1-w\over 1-vw}\nonumber \\
&& \sin\psi_1=-\sin\psi \left({1-v-v w\over 1-v +vw}\right).\nonumber \\
&&\cos\psi=\sqrt{w(1-v)\over 1-v w}
\eeqa

Following Ref.~\cite{ael} we introduce the variable 
\beq
z=-{k_3^\perp\cdot k_2^\perp\over (k_3^\perp)^2},
\label{z}
\eeq
where the perpendicular components are measured in the c.m. system of the 
two incoming partons $p_1$ and $p_2$.
We can covariantize (\ref{z}) in a trivial manner by using light cone 
identities 
\beqa
&& k_3^2=2 k_3^+ k_3^- - (k_3^\perp)^2=0\nonumber \\
&& (k_3^\perp)^2=2 k_3^+k_3^-=2 {k_3\cdot p_1 k_3\cdot p_2\over p_1\cdot p_2}
\eeqa

Similarly, expanding $k_3\cdot k_2$ in its light cone components and 
covariantazing we get 
\beq
k_3^\perp\cdot k_2^\perp=-k_3\cdot k_2 + 
{k_3\cdot p_2 k_2\cdot p_1 + k_3\cdot p_1k_2\cdot p_2\over p_1\cdot p_2}
\eeq
and
\beqa
&& z={s k_3\cdot k_2 +u k_2\cdot p_1 +t k_2\cdot p_2\over t u}\nonumber \\
&& \equiv m \cdot k_2,
\label{zz}
\eeqa
where the four-vector $m$ is defined by 
\beq
m\equiv {s k_3 +t p_2 +u p_1\over t u}.
\label{m}
\eeq
(\ref{zz}) is the covariant expression of $z$. At this point, however, 
it is necessary to evaluate $m$ in the c.m. system of the pair $(1,2)$. 

Using (\ref{kk}) in Eq.~(\ref{m})
it is a simple exercise to show that $m$ has only longitudinal 
components given by 
\beqa
&& m=({s\over t u})^{1/2}(\sqrt{w(1-v)\over 1-w},0,...,0,\sqrt{1-v w\over 1-w})
\nonumber \\
&&=({s\over t u})^{1/2}(m'_0,0,...,0,m'_n).
\eeqa
Notice that $\cos\psi=m'_0/m'_n \equiv \tanh\,\chi$.

In order to isolate a pair $(1,2)$ of a given $z$, we introduce the identity 
\beq
1=\int dz \delta(z-m.k_2)
\eeq
in the expression of the phase space $Ps_3$ which becomes
\beqa
&& Ps_3 ={s^{2-2\epsilon} \pi^{1-2\epsilon}\over 2^4 
(2 \pi)^{5-4\epsilon}
2^8\Gamma [1-2\epsilon]}\left({-\epsilon\over (1-\epsilon)}\right)
\int dv dw (1-v)^{-\epsilon} (1-w)^{1-\epsilon}v^{2-2 \epsilon}
w^{-\epsilon}\nonumber \\
&& \,\,\,\,\,\,\,\,\,\,\,\,\,\,\,\, \times\int_{0}^{\pi}d\theta_2 dz 
\delta(z-m\cdot k_2) \sin^{2-2\epsilon}\theta_2\int_{0}^{\pi}d\theta_1 
{\sin^{3-2\epsilon}\theta_1}.
\eeqa

The integration over $\theta_1$ can now be performed and the remaining 
$\delta$-function eliminated. One gets

\beqa
&&\int_{0}^{\pi}d\theta_1 \sin^{3-2\epsilon}\theta_1\delta(z-m\cdot k_2)
\nonumber \\
&&=\int_{0}^{\pi}d\theta_1\sin^{3-2\epsilon}\theta_1
\delta(z-1/2 +1/2 \cos\theta_1 \coth\,\chi)\nonumber \\
&&=2\,\, \tanh\,\chi g(v,w,z)^{1-\epsilon}
\eeqa
where 
\beq
g(v,w,z)={1-w +4 w (1-v)z(1-z)\over 1-v w}
\label{gi}
\eeq
The final phase space therefore can be cast in the form 
\beqa
&&Ps_3={ \pi^{1-2\epsilon} s^{2-2\epsilon}
\over 2^4 (2\pi)^{5 -4\epsilon}}
\left({-\epsilon\over (1-\epsilon)}
\right) 2\,\, \tanh\,\chi\,\,
 g(v,w,z)^{1-\epsilon}\int dv\,dw\, v^{2-2\epsilon}
w^{-\epsilon}(1-w)^{1-\epsilon}(1-v)^{-\epsilon}\nonumber \\
&&\,\,\,\,\,\,\,\,\,\,\,\times \int_{0}^{\pi}d\theta_2 
\sin^{2-2\epsilon}\theta_2\nonumber \\
\label{ps3}
\eeqa

\section{Evaluation of the Phase Space Integrals}
From the definition of the hypergeometric function 
\beq
F[a,b,c,z]= {2^{1-c}\Gamma[c]\over \Gamma[b]\Gamma[c-b]}
\int_{0}^{\pi}{\sin^{2 b-1}\theta(1+\cos\theta)^{c-2 b}d\theta 
\over (1-z/2+z/2 \cos\theta)^{a}}
\eeq
we easily get
\beq
I[a,2 b-1]\equiv \int_{0}^{\pi} {\sin^{2 b -1}\theta d\,\theta\over (\alpha 
+\beta \cos\theta)^a}
={\Gamma^2[b]\over \alpha^a 2^{1-2 b}\Gamma[2 b]}F[a/2,a/2 +1/2,b+1/2,
\beta^2/\alpha^2]
\eeq
We now use the relation
\beq
F[\alpha,\beta,\gamma,z]=(1-z)^{\gamma-\alpha-\beta}F[\gamma-\alpha,
\gamma-\beta,\gamma,z]
\eeq
to get 
\beq
I[2,2-2 \epsilon]={\pi 2^{2\epsilon-2}\over 
\alpha^{1-2\epsilon}}{\Gamma[3-2\epsilon]\over 
\Gamma^2[2-\epsilon]\,(\alpha^2-\beta^2)^{\epsilon +1/2}}F[1-\epsilon,
1/2-\epsilon, 2-\epsilon,\beta^2/\alpha^2],
\eeq
and
\beq
I[1,-2 \epsilon]={\pi 2^{2\epsilon}\over \alpha^{-2\epsilon}}
{\Gamma[1-2\epsilon]\over 
\Gamma^2[1-\epsilon]\,(\alpha^2-\beta^2)^{\epsilon +1/2}}F[1/2-\epsilon,
-\epsilon, 1-\epsilon,\beta^2/\alpha^2].
\eeq
At the end the result is expressed in terms of ``plus'' distributions using 
various identities whose derivation is briefly discussed in the next appendix. 

For instance, let's denote by $Ps_3[t_3]$ the phase space contribution due to a 
factor $1/t_3$ in the matrix element. 
We get
\beq
Ps_3[t_3]={1\over t_{30}}K_{unp}\left({h^2\over g}\right)^{\epsilon}
{1\over |1-z|^{1 + 2 \epsilon}}
F[1/2-\epsilon,-\epsilon,1-\epsilon, \beta^2/\alpha^2].
\eeq
where the factor $K_{unp}$ is the same as calculated in the unpolarized case
\beq
K_{unp}={s^{1-2\epsilon}(4\pi)^{2\epsilon}2^{2\epsilon-8}\over \pi^3 
\Gamma^2[1-\epsilon]}(1-v)^{-\epsilon}(1-w)^{-\epsilon}v^{1-2 \epsilon}
w^{-\epsilon}
\eeq

Notice that at $z=1$, we have $\beta=\alpha$ and Taylor expanding the 
hypergeometric function around $z=1$ we get (with $r(z)\equiv \beta^2/\alpha^2$)
\beqa
&& F[a,b,c, r(z)] =
F[a,b, c,1] +{a\,b\,\over  c} F[a+1,b+1,c+1,1] r'(1)
\nonumber \\
&& = F[a,b,c,1] + (z-1)O(\epsilon).
\eeqa
Then we can set $F[a,b,c,r(z)]=F[a,b,c,1] + O(\epsilon)$ with 
$F[a,b,c,1]=1/2^{2 \epsilon} + O(\epsilon^2)$.
We finally get (see appendix C)
\beq
Ps_3(1/t_3)={1\over t_{30}}K_{unp}\left({\theta(1-z)\over (1-z)_+} 
+{\theta(z-1)\over (z-1)_+}
+\delta(z-1)(-{1\over \epsilon}- \log \,z_{max})\right).
\eeq
As an example of applications of of the methods discussed above in the polarized 
case, let's consider the contribution to the final phase space coming from 
matrix elements of the form ${\widehat{\widehat{ k}}}^2/t_3^2$. We get

\beqa 
&& Ps_3\left({\widehat{\widehat {k}}^2\over t_3^2}\right)={1\over t_{30}^2}
 K_{pol}
\left( {-\epsilon\over 2(1-\epsilon)}\right)
\left( {h^2\over g}\right)^\epsilon {1\over |1-z|^{1+2 \epsilon}}\eta(z)g(z)
F[1-\epsilon,1/2-\epsilon,2-\epsilon, \beta^2/\alpha^2]\nonumber \\
\eeqa

where
\beq
\eta(z)={1\over(1 + \tanh^2\chi(1-2 z))}
\eeq 
and
\beqa
&&K_{pol}={\pi^{2-2\epsilon} s^{2-2\epsilon}2^{2\epsilon}\over 2^6 
(2 \pi)^{5-4\epsilon}}{\Gamma[3-2\epsilon]\over \Gamma[2-\epsilon]^2}
 v^{2-2\epsilon}
w^{-\epsilon}(1-w)^{1-\epsilon}(1-v)^{-\epsilon}\nonumber \\
\eeqa
For future purposes it is convenient to introduce the function 
$\sigma(z)\equiv \eta(z) g(z)$.
We get 

\beqa
&&Ps_3\left({\widehat{\widehat {k}}^2\over t_3^2}\right)
={1\over t_{30}^2}K_{pol}\left({-2\over 1-\epsilon}\right)\delta(z-1)
\nonumber \\
\label{uno}
\eeqa
\beqa
&& Ps_3\left({\widehat{\widehat {k}}^2\over u_3^2}\right)
={1\over u_{30}^2}K_{pol}\left({-2\over 1-\epsilon}\right)\delta(z-1)
\label{due}
\eeqa
Notice that in the unpolarized case - as discussed in Ref.~\cite{ael} - 
in some specific matrix elements, such as $Ps_3\left(1/(s_{13}u_3)\right)$,
 singularities in both variables $v$ and $w$ are encountered (for $w=1$ and 
$z=1$). 
The regularization in terms of plus-functions of the corresponding contributions 
and a detailed 
discussion of the derivation, which is similar 
to the polarized case can be found in Ref.~\cite{ael}.

\section{Identities for ``Plus'' Distribution}
As we have discussed before, we take $z$ to be positive from the beginning. 

In the expression of the final cross section, we are going to encounter 
singularities at $z=1$ (and at $w=1$) which have to be regulated 
in an appropriate way by the use of ``plus'' distribution. 
Identities for ``plus'' distribution can be easily derived by the 
integration by parts method, as we are going to illustrate briefly
by an example. Assuming $z>0$, we define two kinds of regulated 
``plus' distributions $\theta(1-z)/(1-z)_+$ and $\theta(z-1)/(z-1)_+$ by
\beqa
&& \int_{0}^{1}{f[z]\theta(1-z)\over (1-z)_+}dz\equiv \int_{0}^{1}dz 
{f[z]-f[1]\over 1-z}\nonumber \\
&& \int_{1}^{z_{max}}{f[z]\theta(z-1)\over (z-1)_+}
\equiv \int_{1}^{z_{max}}{f[z]-f[1]\over (1-z)},
\eeqa

Let's define the two functions 
\beqa
&& h(z)={1 + \tanh^2\chi (1-2 z)\over 2 \tanh\chi}\nonumber \\
&& g(z)={1-w +4 z w )(1-v)(1-z)\over 1-v w}
\eeqa
with $h(z_{max})=-z_{min}$. 
Then we get the identities 
\beqa
&& \int_{0}^{z_{max}}dz f[z]{\theta(z-1)\over (z-1)^{1+2\epsilon}}
\left({h^2[z]\over g[z]}\right)^{\epsilon}
=\int_{0}^{z_{max}}dz f[z]\left[{\theta(z-1)\over (z-1)^{1+2\epsilon}}
\left({h^2[z]\over g[z]}\right)^{\epsilon}\right]_+ \nonumber \\
&&\,\,\,\,\,\,\,\,\, +f[1]\left( -{1\over 2 \epsilon} +\log(z_{max}-1)
-{1\over 2}
\log({h^2[1]\over g^2[1]})\right). 
\eeqa
We have integrated once by parts 
and the boundary (regular) terms have been set to zero before taking 
the limit $\epsilon\to 0$ and expanding in $\epsilon$. 
The procedure has to be repeated once more if the boundary terms are still 
singular at the edge of the interval of integration. 
Now we use $z_{max}-1=-z_{min}$ together with the simple identity 
\beq
\theta(z-1)\left[{\left(h^2[z]\over g[z]\right)^{\epsilon}}{1\over (z-1)^{1+
2\epsilon}}\right]_+=\theta[z-1]{1\over (z-1)_+} +O[\epsilon]
\eeq
to finally obtain the relation 
\begin{eqnarray}
\theta(z-1)\left[{\left(h^2[z]\over g[z]\right)^{\epsilon}}{1\over (z-1)^{1+
2\epsilon}}\right]&=&\theta[z-1]{1\over (z-1)_+}
+\delta(z-1)\nonumber \\
& &\left(-{1\over 2 \epsilon}+\log(-z_{min}) -{1\over 2}
\log({h^2[1]\over g^2[1]})\right).
\label{unop}
\end{eqnarray}
In order to derive a similar identity in the case of  $0<z<1$ we 
proceed in a similar way 

\beqa
&& \int_{0}^{1}dz f[z]{\theta(1-z)\over (z-1)^{1+2\epsilon}}
\left({h^2[z]\over g[z]}\right)^{\epsilon}
=\int_{0}^{1}dz f[z]\left[{\theta(z-1)\over (z-1)^{1+2\epsilon}}
\left({h^2[z]\over g[z]}\right)^{\epsilon}\right]_+ \nonumber \\
&&\,\,\,\,\,\,\,\,\, +f[1]\left( -{1\over 2 \epsilon} -{1\over 2}
\log({h^2[1]\over g^2[1]})\right). 
\eeqa
Further manipulations similar to those presented above then give
\beq
\theta[1-z]\left[{\left(h^2[z]\over g[z]\right)^{\epsilon}}{1\over (z-1)^{1+
2\epsilon}}\right]=\theta[1-z]{1\over (z-1)_+}
+\delta(1-z)\left(-{1\over 2 \epsilon}-{1\over 2}\log({h^2[1]\over g[1]})
\right).
\label{duep}
\eeq

Combining (\ref{unop}) and (\ref{duep}), after some manipulations we get the 
identity 
\beq
 {1\over |1-z|^{1+2\epsilon}}\left({h^2[z]\over g[z]}\right) 
={\theta(1-z)\over (1-z)_+} + {\theta(z-1)\over (z-1)_+}
+\delta(1-z)(-{1\over \epsilon} -\log z_{max}).
\eeq

In a similar way we can derive the identity
\beqa
&& {1\over |z-1|^{1+ 2\epsilon}}\left({h^2(z)\over g(z)}\right)^{\epsilon}
\eta(z)g(z)F[z]= 
\theta(1-z)\left({1\over 1-z}\eta(z)g(z)\right)_+ \nonumber \\
&& + \theta(z-1)\left({1\over z-1}\eta(z)g(z)\right)_+
-{1\over \epsilon}\delta(z-1)\,\eta(1)\, g(1)\,F[1] + \,\,\, O(1)
\eeqa
Using $F[1]= 2(1+ O(\epsilon))$
where $F[z]\equiv F[1-\epsilon,1/2-\epsilon,2-\epsilon,r(z)]$ we get 
(\ref{uno}) and (\ref{due}).

\pagebreak

\noindent
{\bf Figure Captions}

\newcounter{num}
\begin{list}%
{[\arabic{num}]}{\usecounter{num}
    \setlength{\rightmargin}{\leftmargin}}

\item (a) Lowest order Feynman diagrams for double prompt photon
production. (b) Examples of single fragmentation contributions to double
prompt photon production. (c) Example of box diagram contribution to
double prompt photon production.

\item (a) Examples of virtual corrections to the lowest order diagrams.
(b) Examples of next-to-leading order three-body final-state diagrams for 
the $q \bar{q}$ initial state.

\item Examples of contributions to the higher order $qg$ initiated
process.

\item Unpolarized non-isolated cross section $d\sigma/dk^\gamma_{T1} dy_1dz$ 
as a function of $z$
for $p + p\rightarrow \gamma + \gamma + X$ at $\sqrt{s}=500$ GeV.  We set 
$y_1 = 0$.  Results are presented in the form of a histogram in bins of
width $\Delta z=0.2$.  In (a), for $k^{\gamma}_{T1}=5$ GeV, we show the net 
contribution from the lowest order process $q \bar{q} \rightarrow \gamma
\gamma$  and from all the higher order processes and the full sum.
In (b) we compare the polarized and unpolarized cross sections for the
same parameters as in (a). The unpolarized has been multiplied by 0.1
for easier comparison.

\item (a) The transverse momentum dependence of 
$d\sigma/dk^\gamma_{T1} dy_1 dz$, the non-isolated cross section, for $z$ 
integrated over the 
interval $0.2 <z <2.0$.  The upper solid line shows the full unpolarized
cross section. The lower curves show the polarized cross section as
given by the two different scenarios for the polarized parton
distributions discussed in the text. 
(b) The longitudinal asymmetry, defined in the text, for the
non-isolated cross section as predicted by parton distributions
assuming scenarios a and b. The asymmetry for the lowest order Born
cross section is included for comparison.

\item (a) $k_T$ dependence of the isolated cross section integrated over
rapidity range $-3\leq y \leq 3$ using isolation parameters given in the
text. The unpolarized cross section as well as the polarized cross
section assuming scenarios a and b. (b) The longitudinal asymmetry for
the cross section in (a) as predicted by scenario a and b. (c) The same
asymmetry as in (b), but only for scenario a broken down into
contributions from the $q\bar{q}$, $qg$ and $gg$ initiated processes.

\item The renormalization/factorization scale $\mu$ dependence.  For the
sum of all contributing subprocesses, 
$d\sigma/dk^\gamma_{T1}$, for $-3 \leq y\leq 3$ 
is shown as a function of $k^\gamma_{T1}$
for three values of $\mu=n( (k^\gamma_{T1})^2+(k^{\gamma}_{T2})^2)/2$: 
0.5, 1.0, and 2.

\item The same cross section as in fig.6a for the unpolarized case but with
$\sqrt{S}=14$ TeV.

\end{list}


\begin{figure}
\centerline{\psfig{file=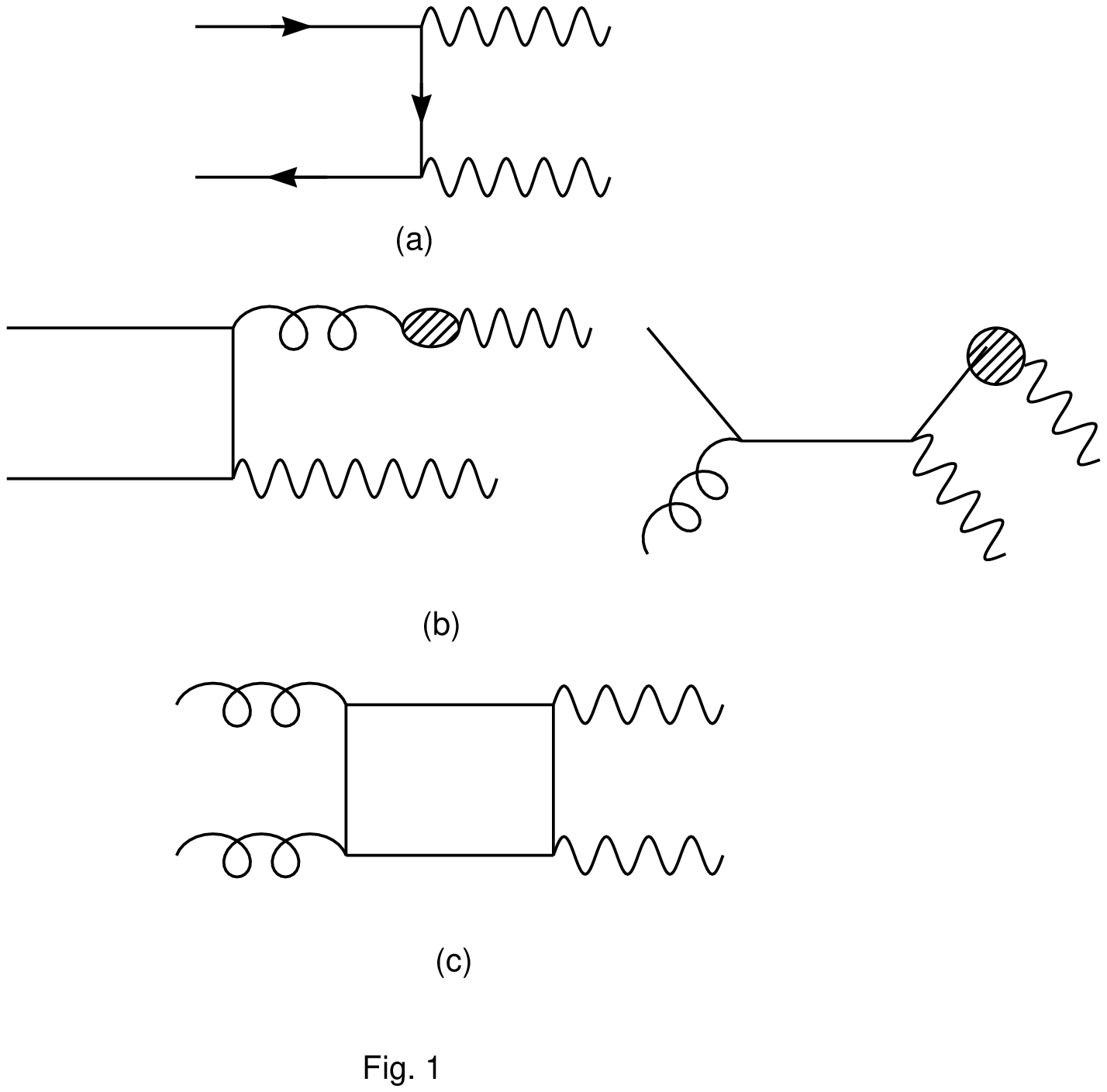,height=14cm}}
\end{figure}

\begin{figure}
\centerline{\psfig{file=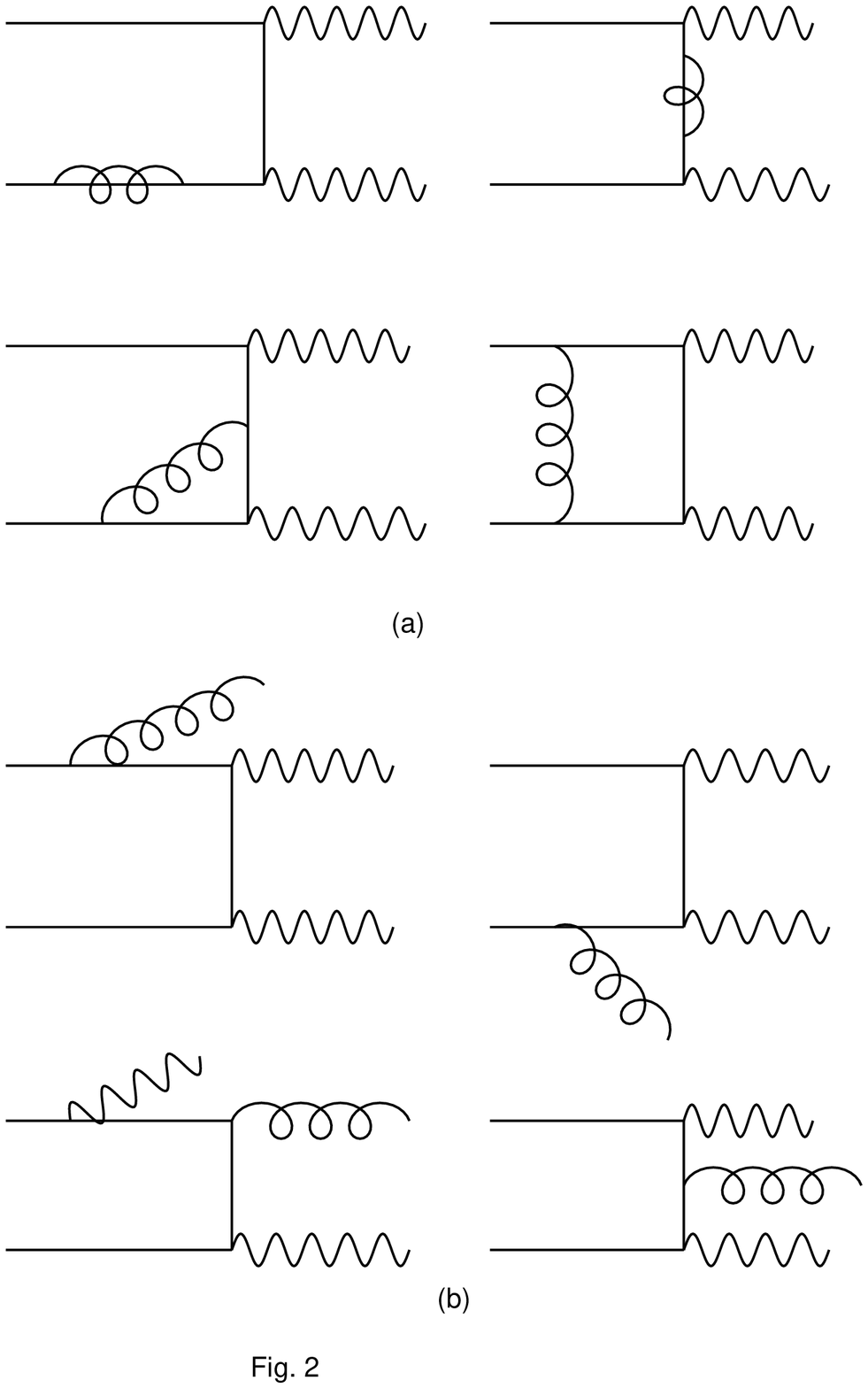,height=12cm}}
\end{figure}

\begin{figure}
\centerline{\psfig{file=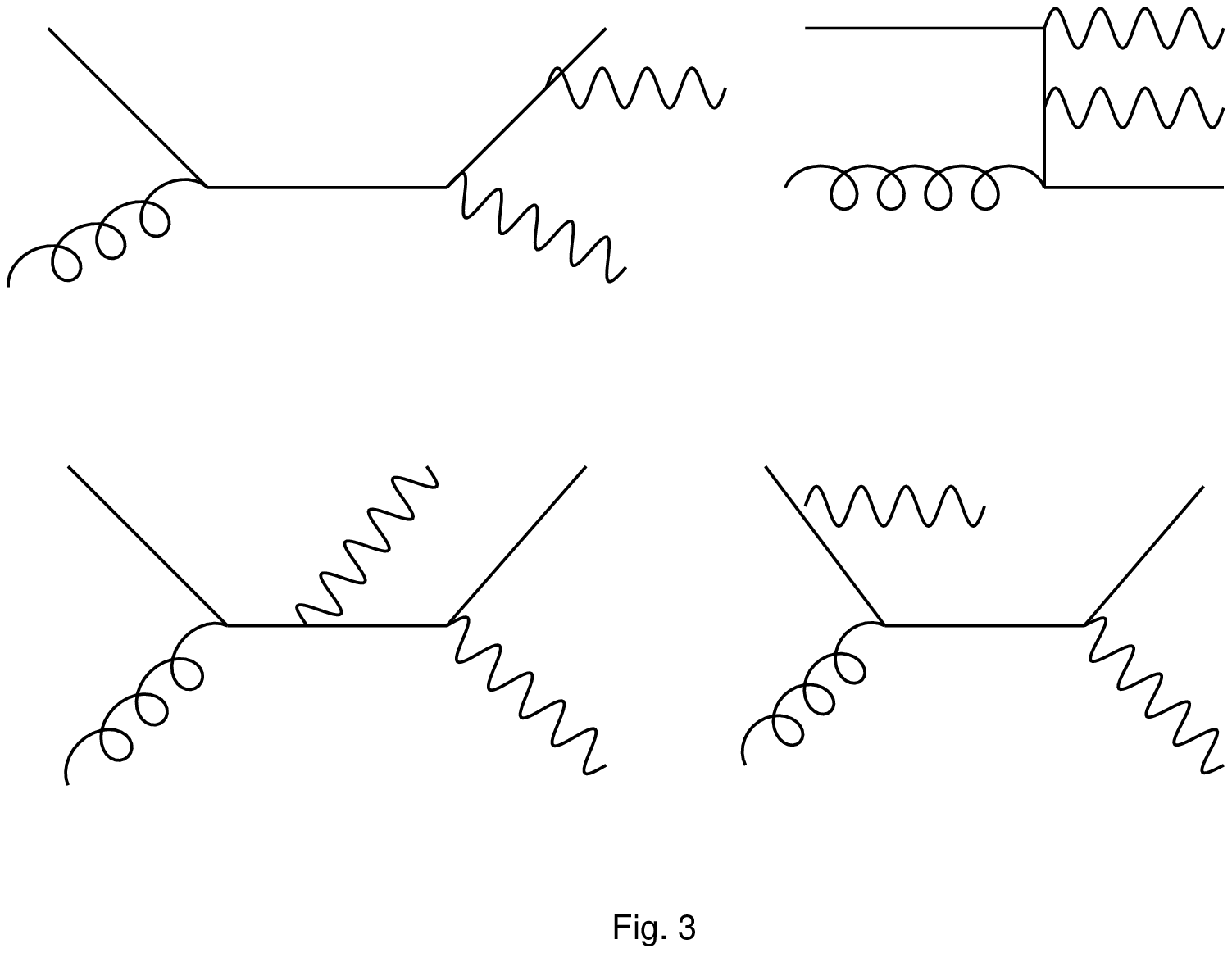,height=8cm}}
\end{figure}

\epsffile{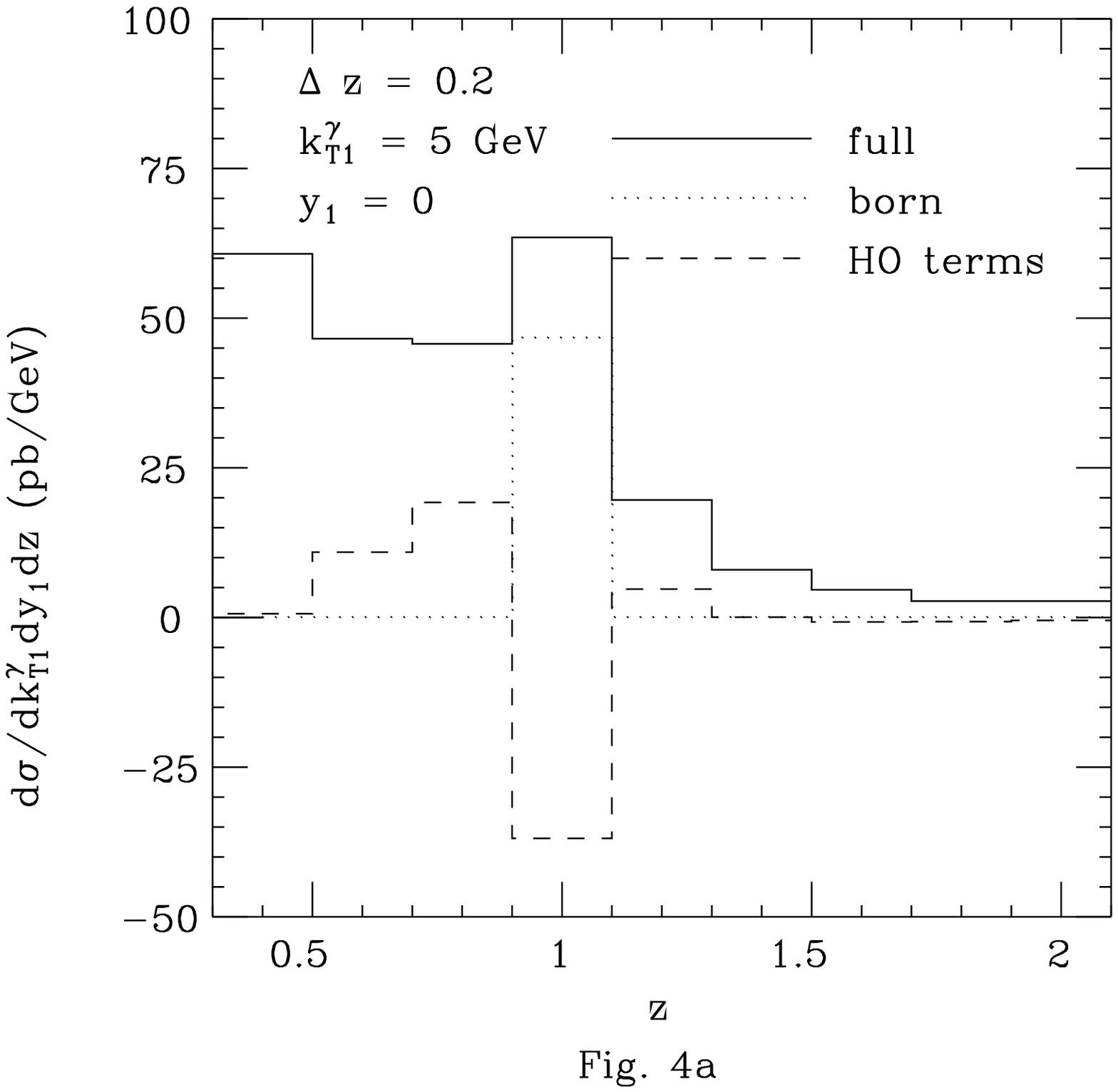}

\epsffile{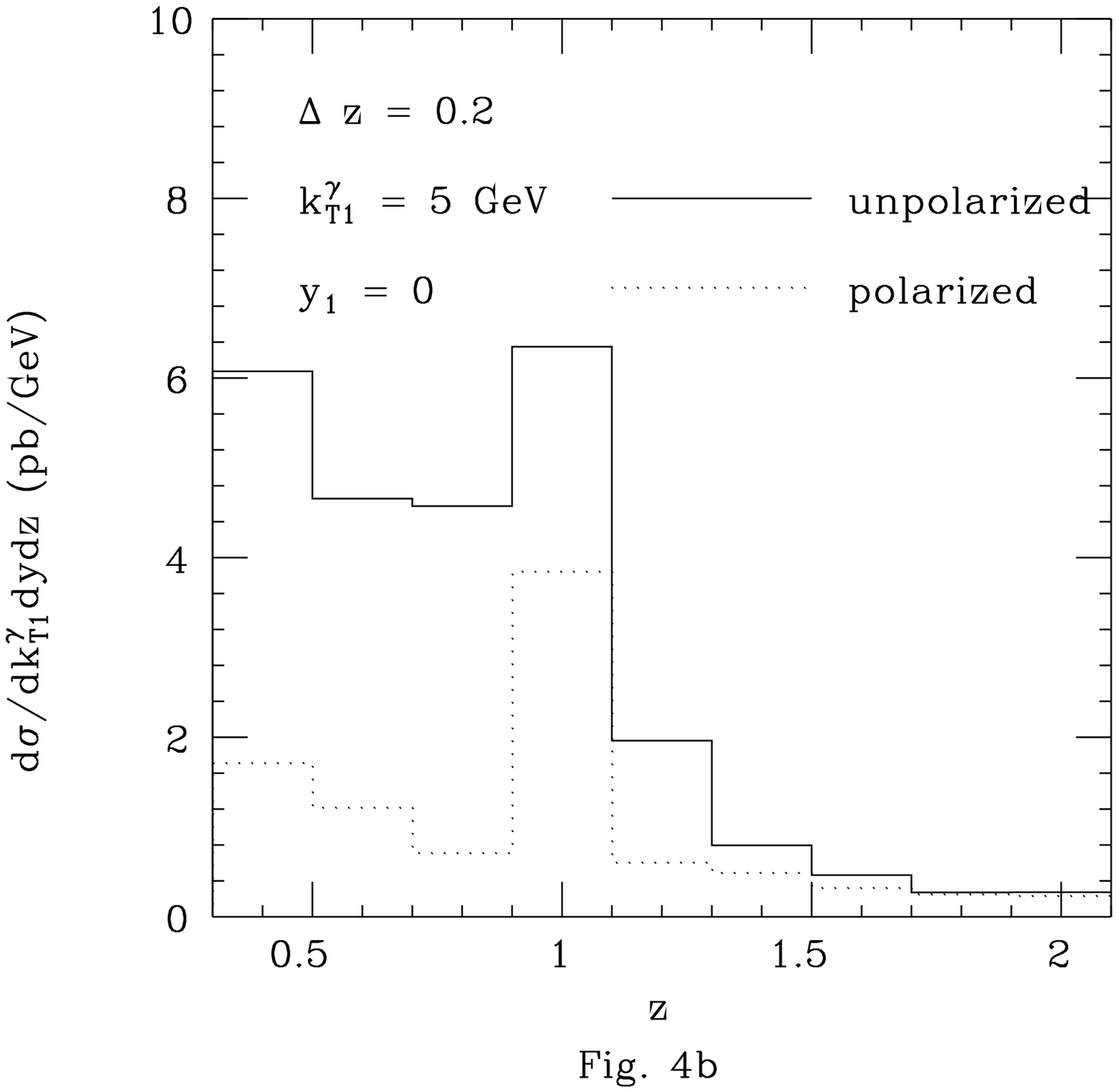}

\epsffile{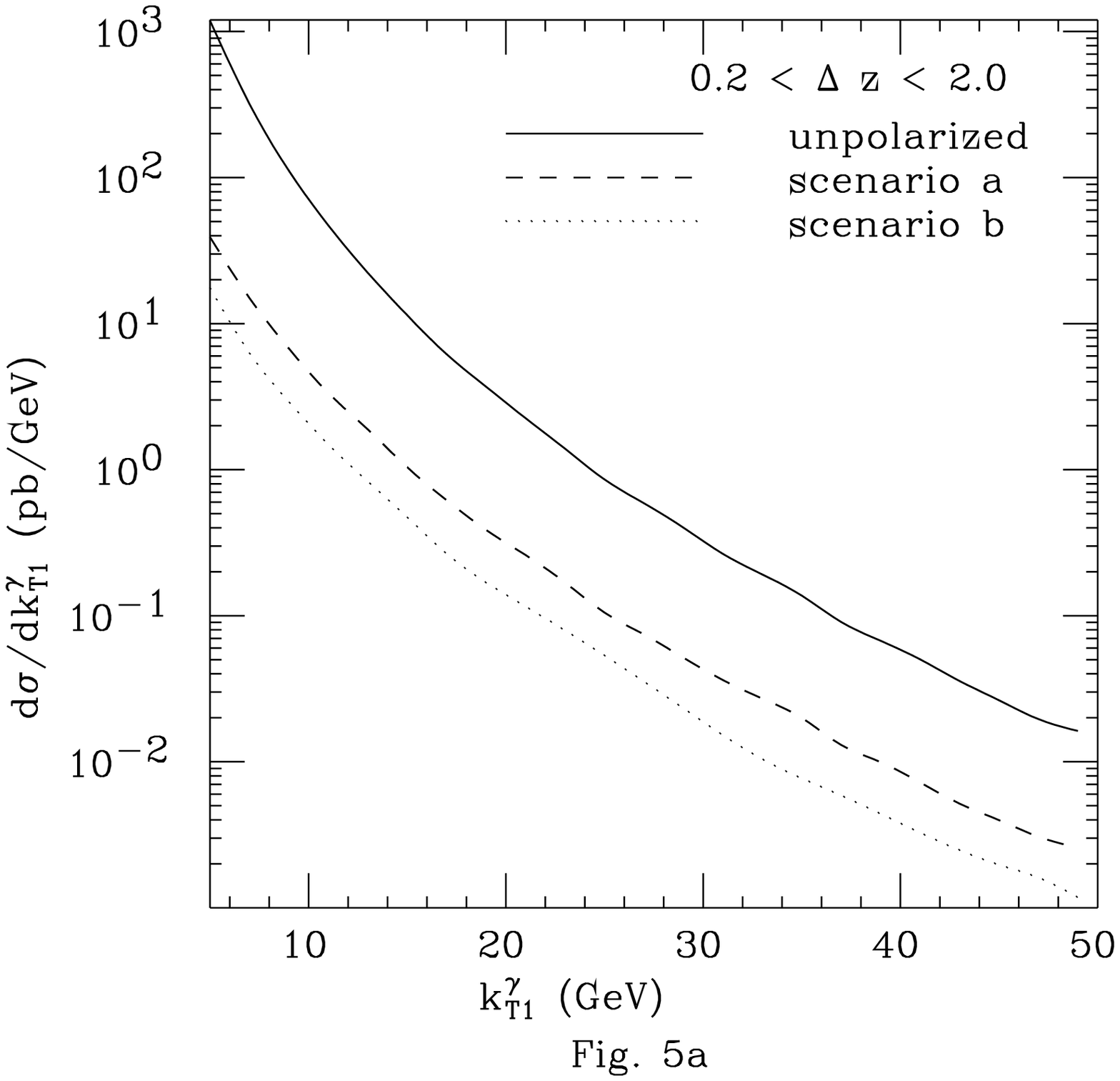}

\epsffile{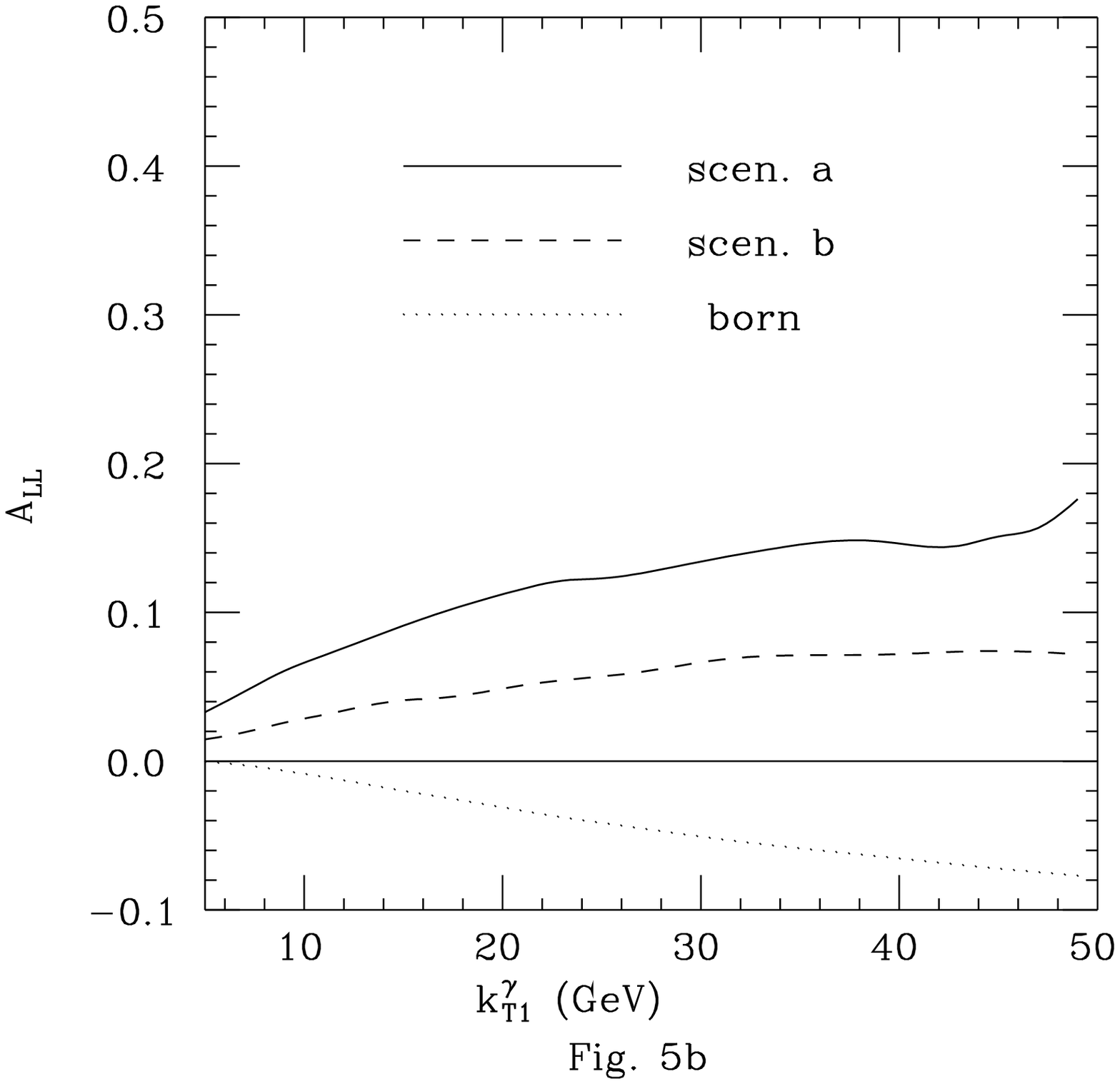}

\epsffile{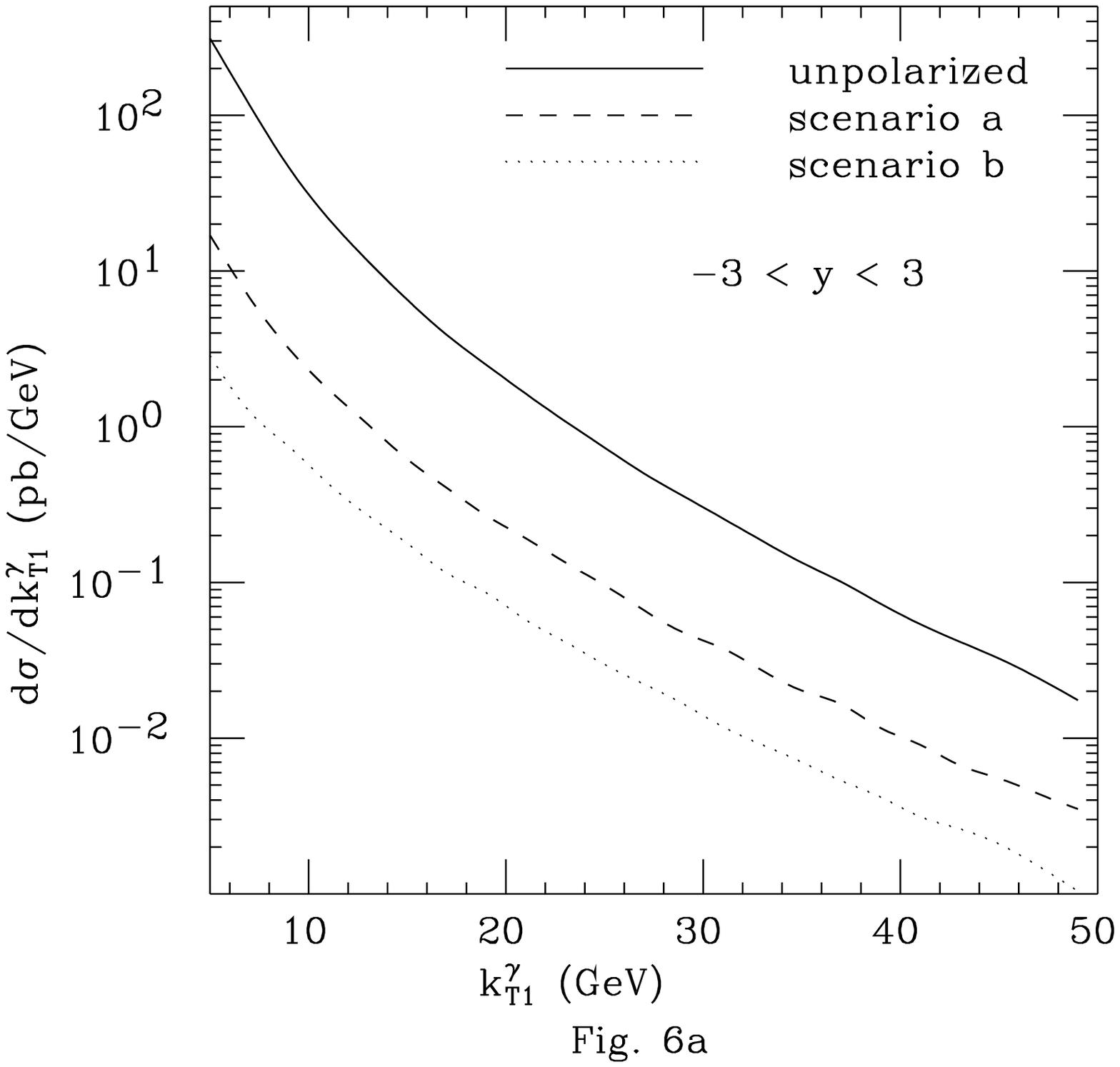}

\epsffile{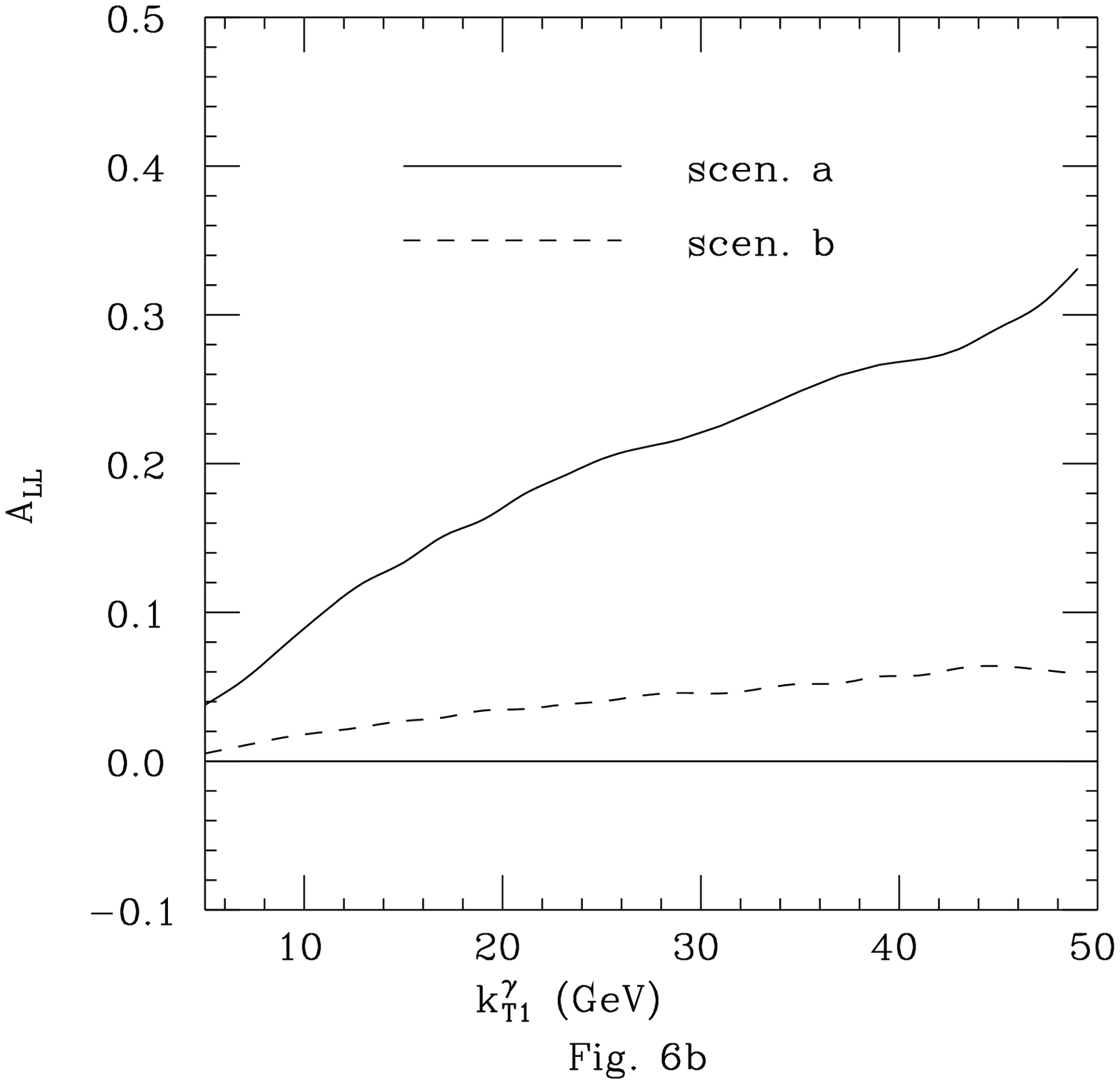}

\epsffile{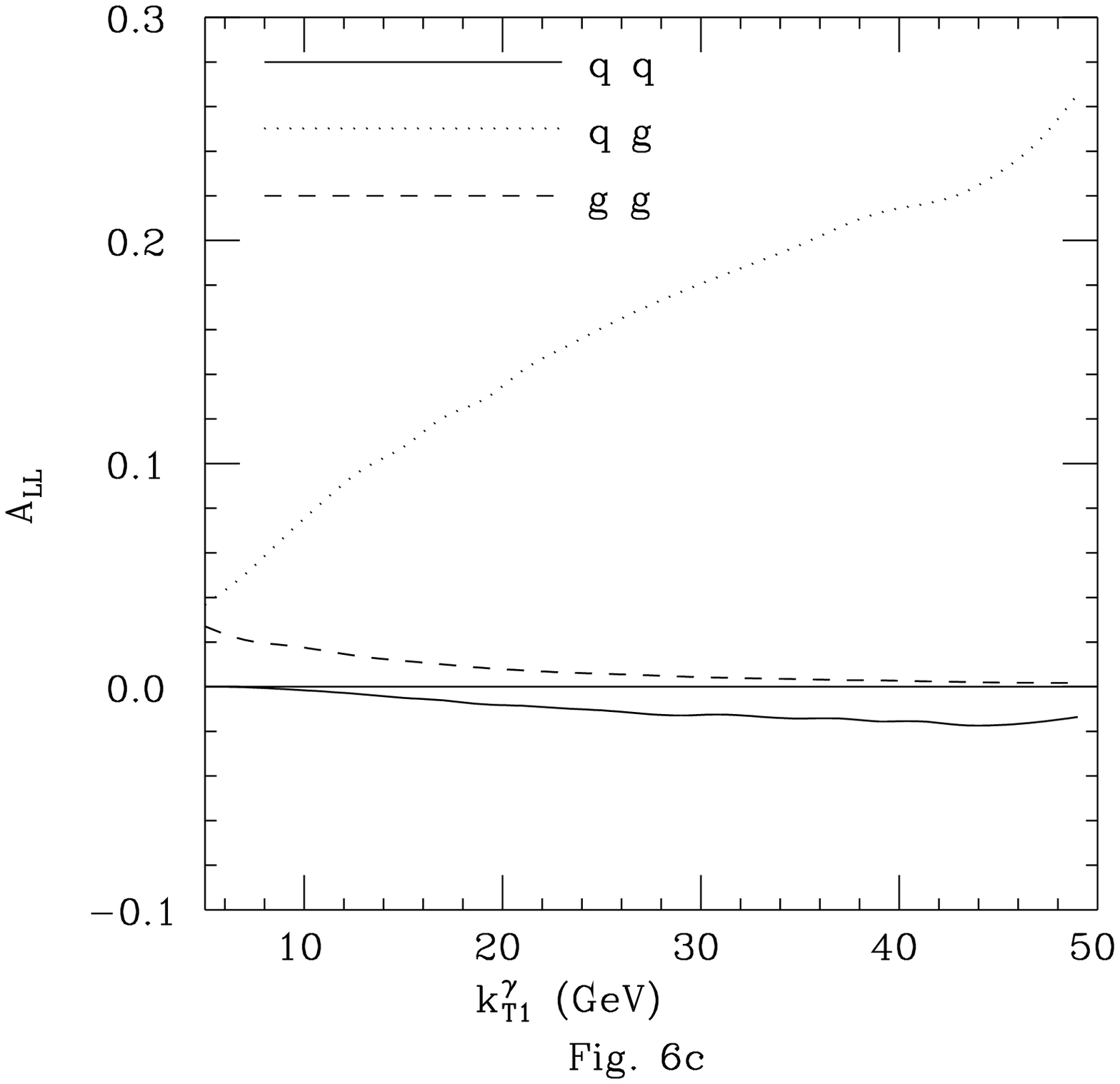}

\epsffile{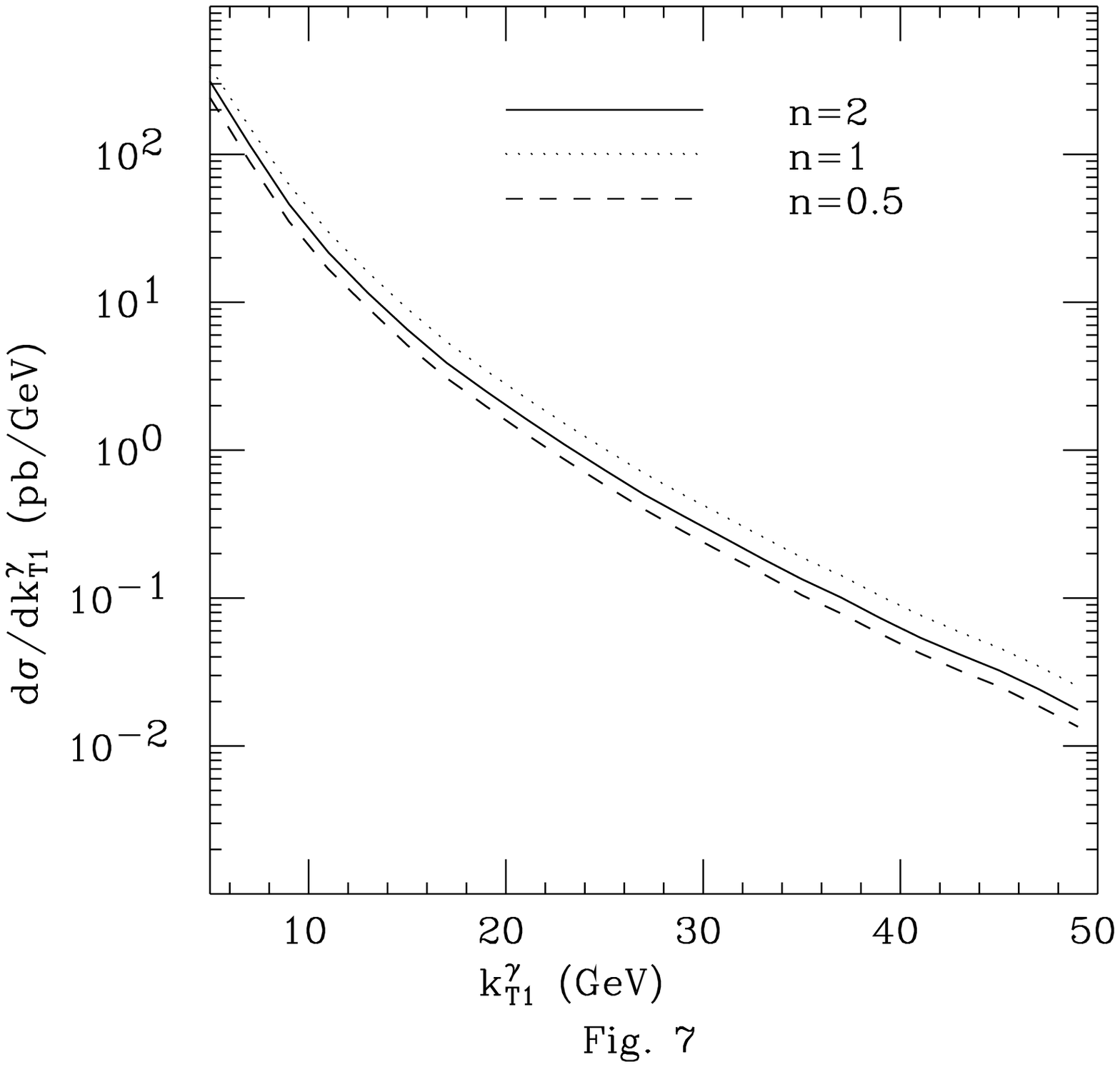}

\epsffile{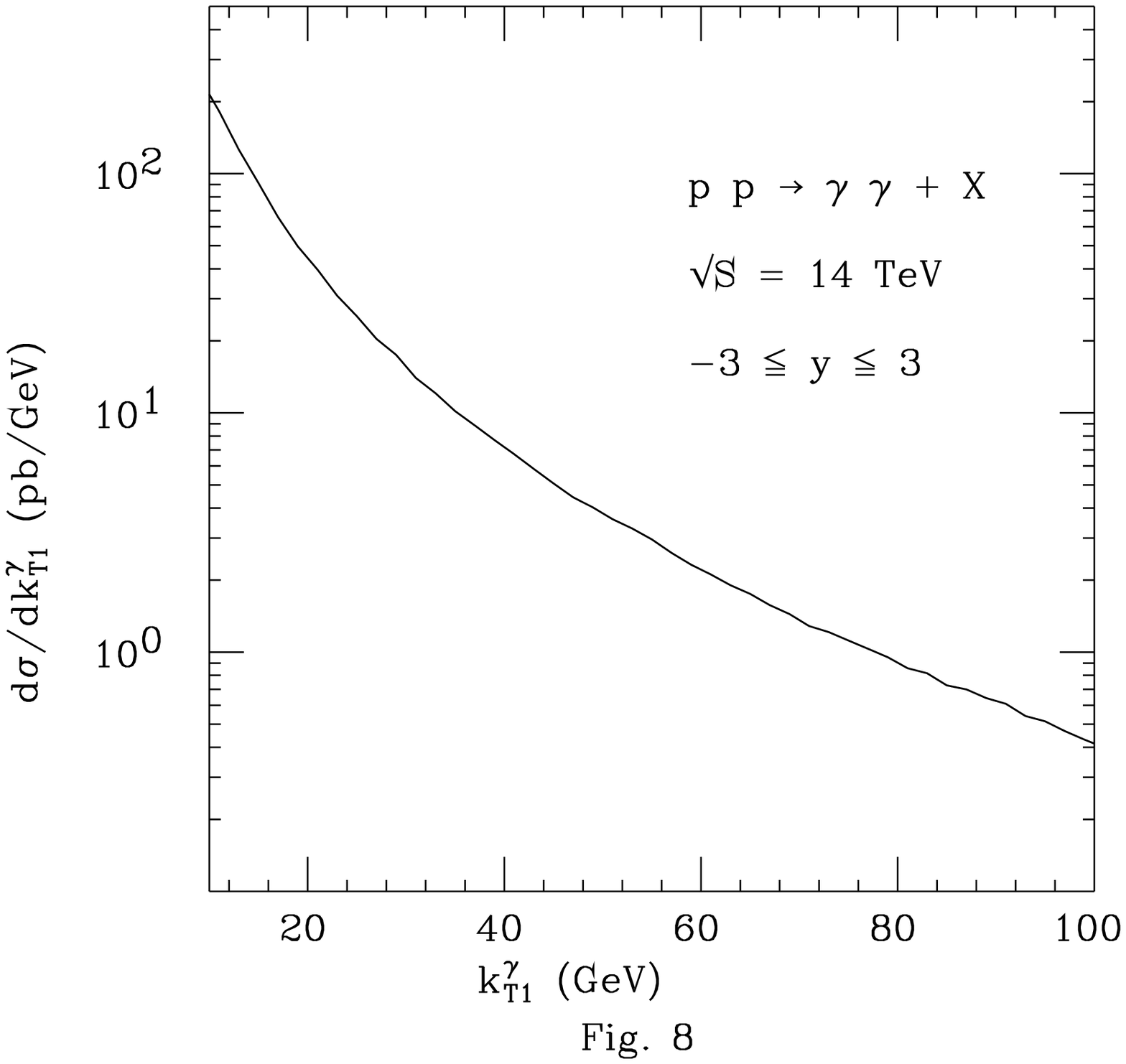}


\begin{thebibliography}{99}
\bibitem{EMC} J. Ashman et al., EMC collaboration, Phys. Lett. {\bf
B206} (1988) 364,
 Nucl. Phys. {\bf B328} (1989) 1.
\bibitem{set1} A. V. Efremov and O. V. Teryaev, 
Czech. Hadron Symposium (1988) 302; 
G. Altarelli and G. G Ross, 
Phys. Lett. {\bf B212} (1988)391; 
R. D. Carlitz, J. C. Collins and A. H. Mueller, Phys. Lett. {\bf B214} (1988) 
229;  G. Bodwin and J. Qiu, Phys. Rev. {\bf D41} (1990) 391. 

\bibitem{set2} SMC Collaboration, B Adeva et al. Phys. Let. {\bf B302} (1993) 
533, ibidem {\bf B320} (1994)400;
 SMC Collaboration, D. Adams et al., Phys. Lett. {\bf B329} (1994) 399;
E143 Collaboration, K. Abe et al., SLAC-PUB-6508 (1994) preprint;
E143 Collaboration, R. Arnold et al., presented at ICHEP94, 
Glasgow, August 1994.  
\bibitem{EJ} J. Ellis and R. Jaffe, Phys. Rev. {\bf D9} (1974) 1444; Erratum 
{\bf D10} (1974) 1669. 
(1989) 307.
\bibitem{ans} M. Anselmino, A. Efremov and E. Leader, Phys Rep. 261,
NO.'s {\bf 1, 2}.
\bibitem{vn} E. B. Zijlstra and W. L. Van Neerven, Nucl. Phys. {\bf B417} 
(1994)
61, Erratum {\bf B245} (1994) 245. R. Mertig and W. L. Van Neerven 
NIKHEF-H/95-031. W. Vogelsang, Rutherford Preprint, RAL-TR-95-071.
\bibitem{robinett} R. W. Robinett, ANL-HEP-CP-95-28, To appear in the
proceedings of The International Symposium on Particle Theory and
Phenomenology, Iowa State University, May 22-24, 1995.
\bibitem{nlo1} A. P. Contogouris, B. Kamal, Z. Merebashvili and F. V.
Tkachov, Phys. Lett. {\bf B304} 329 (1993); Phys. Rev. {\bf D48} 4092
(1993). P. Ratcliffe, Nucl. Phys. {\bf B223} 45, (1983).
\bibitem{nlo2} L. E. Gordon and W. Vogelsang Phys. Rev. {\bf D48} (1993) 3136.
 \bibitem{ael} P. Aurenche et al. Z. Phys. {\bf C29}, (1985) 459. B. Bailey,
J. Ohnemus and J. F. Owens Phys. Rev. {\bf D46}, 2018 (1992).
\bibitem{rd} M. A. Doncheski and R. W. Robinett, Phys. Rev. {\bf D46},
2011 (1992)  
\bibitem{cg} C. Corian\`{o} and L. E. Gordon, to appear. While
completing this work we found that new distributions for the polarized
proton have been completed by Gluck et al., and Gehrmann and Stirling.
\bibitem{owens} J. F. Owens, Rev. Mod. Phys. {\bf 59}, 465 (1987).
\bibitem{thv} G. t'Hooft and M. Veltman, Nucl. Phys. B44, 189 (1972).
\bibitem{bm} P. Breitenlohner and D. Maison, Comm. Math. Phys. {\bf 52}, 
11 (1977).
\bibitem{alt} G. Altarelli and G. Parisi, Nucl. Phys. {\bf B126}, 298
(1977).
\bibitem{grv} m. Gl\"{u}ck, E. Reya and A. Vogt, Phys. Rev. {\bf D45},
3986 (1992).
\bibitem{chengwai} H.-Y. Cheng and C. F. Wai, Phys. Rev. {\bf D46}, 125
(1992).
\bibitem{bbf} E. L. Berger, E. Braaten and R. D. Field, Nucl. Phys. {\bf
B239}, 
(1984) 52. 
\bibitem{ellis} R. K. Ellis, M. A. Furman, H. E. Haber and I. Hinchliffe, 
Nucl. Phys {\bf B173} (1980) 397.
\bibitem{bonn} G. Bonneau, Phys. Lett. {\bf 96B} 147 (1980). 
\end{thebibliography}
\end{document}